\documentclass[conference]{IEEEtran}
\IEEEoverridecommandlockouts
\usepackage{amsmath,amssymb,amsfonts,amsthm}
\usepackage{textcomp}
\usepackage{xcolor}
\usepackage{cite}
\usepackage{bbm}
\usepackage{bm}
\usepackage[hidelinks]{hyperref}
\usepackage{graphicx}

\usepackage{algorithm}
\usepackage{algpseudocode} 

\usepackage[acronym,shortcuts]{glossaries}
\usepackage{tcolorbox}
\definecolor{mycolor}{RGB}{242,242,242}

\newacronym{5G}{5G}{fifth generation}
\newacronym{6G}{6G}{sixth generation}
\newacronym{ADR}{ADR}{antenna decentralized rate}
\newacronym{AER}{AER}{activity error rate}
\newacronym{AP}{AP}{access point}
\newacronym{ADC}{ADC}{analog-to-digital converter}
\newacronym{ABP}{ABP}{approximate belief propagation}
\newacronym{ADD}{ADD}{annealed discrete denoiser}
\newacronym{AMP}{AMP}{approximate message passing}
\newacronym{AUD}{AUD}{active user detection}
\newacronym{AWGN}{AWGN}{additive white Gaussian noise}
\newacronym{BMMSE}{BMMSE}{Bussgang minimum mean square error}
\newacronym{BC}{BC}{belief combining}
\newacronym{BER}{BER}{bit error rate}
\newacronym{BiGAMP}{BiGAMP}{bilinear generalized approximate message passing}
\newacronym{BIP}{BIP}{bilinear inference problem}
\newacronym{GAMP}{GAMP}{generalized AMP}
\newacronym{GF}{GF}{grant-free}
\newacronym{BiGaBP}{BiGaBP}{bilinear Gaussian belief propagation}
\newacronym{BP}{BP}{belief propagation}
\newacronym{BS}{BS}{base station}
\newacronym{CAP}{CAP}{central AP}
\newacronym{CAMP}{CAMP}{convolutional AMP}
\newacronym{CCU}{CCU}{central computing unit}
\newacronym{CDF}{CDF}{cumulative distribution function}
\newacronym{CLT}{CLT}{central limit theorem}
\newacronym{CPU}{CPU}{central processing unit}
\newacronym{CE}{CE}{channel estimation}
\newacronym{CF-mMIMO}{CF-mMIMO}{cell-free massive MIMO}
\newacronym{CG}{CG}{conjugate gradient}
\newacronym{CSI}{CSI}{channel state information}
\newacronym{CSIDCO}{CSIDCO}{complex SIDCO}
\newacronym{DCC}{DCC}{dynamic cooperation clustering}
\newacronym{DFT}{DFT}{discrete Fourier transform}
\newacronym{DoF}{DoF}{degrees of freedom}
\newacronym{DQ}{DQ}{De-quantization}
\newacronym{DL}{DL}{deep learning}
\newacronym{DU}{DU}{deep unfolding}
\newacronym{eMBB}{eMBB}{enhanced mobile broadband}
\newacronym{ECF}{ECF}{estimate-compress-forward}
\newacronym{EP}{EP}{expectation propagation}
\newacronym{EPA}{EPA}{approximate EP}
\newacronym{FA}{FA}{false alarm}
\newacronym{FN}{FN}{factor node}
\newacronym{FG}{FG}{factor graph}
\newacronym{FTN}{FTN}{Faster-than-Nyquist}
\newacronym{GaBP}{GaBP}{Gaussian belief propagation}
\newacronym{IC}{IC}{interference cancellation}
\newacronym{IDD}{IDD}{iterative detection and decoding}
\newacronym{i.i.d.}{i.i.d.}{independent and identically distributed}
\newacronym{JACDE}{JACDE}{joint activity, channel and data estimation}
\newacronym{JACE}{JACE}{joint activity and channel estimation}
\newacronym{JCDE}{JCDE}{joint channel and data estimation}
\newacronym{KLD}{KLD}{Kullback-Leibler divergence}
\newacronym{LE}{LE}{linear estimator}
\newacronym{LSA}{LSA}{latent semantic analysis}
\newacronym{LMMSE}{LMMSE}{linear MMSE}
\newacronym{MAC}{MAC}{multiple-access channel}
\newacronym{MAP}{MAP}{maximum \textit{a posteriori} probability}
\newacronym{MAMP}{MAMP}{memory AMP}
\newacronym{MPDQ}{MPDQ}{message passing DQ}
\newacronym{MD}{MD}{miss-detection}
\newacronym{MF}{MF}{matched filter}
\newacronym{MFB}{MFB}{matched filter bound}
\newacronym{MM}{MM}{moment matching}
\newacronym{MNS}{MNS}{minimum norm solution}
\newacronym{MIMO}{MIMO}{multi-input multi-output}
\newacronym{MU-MIMO}{MU-MIMO}{multi-user MIMO}
\newacronym{MU}{MU}{multi-user}
\newacronym{mMIMO}{mMIMO}{Massive multiple-input multiple-output}
\newacronym{ML}{ML}{maximum likelihood}
\newacronym{MMSE}{MMSE}{minimum mean square error}
\newacronym{mMTC}{mMTC}{massive machine type communications}
\newacronym{MMV-AMP}{MMV-AMP}{multiple measurement vector approximate message passing}
\newacronym{MSE}{MSE}{mean square error}
\newacronym{MUD}{MUD}{multi-user detection}
\newacronym{NLE}{NLE}{nonlinear estimator}
\newacronym{NR}{NR}{new radio}
\newacronym{NMSE}{NMSE}{normalized mean square error}
\newacronym{OAMP}{OAMP}{orthogonal AMP}
\newacronym{OFDM}{OFDM}{orthogonal frequency-division multiplexing}
\newacronym{PDA}{PDA}{probabilistic data association}
\newacronym{PDF}{PDF}{probability density function}
\newacronym{PMF}{PMF}{probability mass function}
\newacronym{PPP}{PPP}{Poisson point process}
\newacronym{PSK}{PSK}{phase-shift keying}
\newacronym{QP}{QP}{quadratic program}
\newacronym{QPSK}{QPSK}{quadrature PSK}
\newacronym{QAM}{QAM}{quadrature amplitude modulation}
\newacronym{SIDCO}{SIDCO}{sequential iterative decorrelation via convex optimization}
\newacronym{SD}{SD}{sphere decoding}
\newacronym{SE}{SE}{state evolution}
\newacronym{SGA}{SGA}{scalar Gaussian approximation}
\newacronym{SIC}{SIC}{soft interference cancellation}
\newacronym{SID}{SID}{self-iterative detection}
\newacronym{SNR}{SNR}{signal-to-noise ratio}
\newacronym{Soft IC}{Soft IC}{soft interference cancellation}
\newacronym{SotA}{SotA}{state-of-the-art}
\newacronym{SVD}{SVD}{singular value decomposition}
\newacronym{SPA}{SPA}{sum-product algorithm}
\newacronym{TX}{TX}{transmit}
\newacronym{RX}{RX}{receive}
\newacronym{UAMP}{UAMP}{unitary AMP}
\newacronym{UE}{UE}{user equipment}
\newacronym{URA}{URA}{unsourced random access}
\newacronym{URLLC}{URLLC}{ultra reliable low latency communications}
\newacronym{VAMP}{VAMP}{vector AMP}
\newacronym{VGA}{VGA}{vector Gaussian approximation}
\newacronym{VN}{VN}{variable node}
\newacronym{w.r.t.}{w.r.t.}{with respect to}
\newacronym{WS}{WS}{warm-started}
\newacronym{ZF}{ZF}{zero-forcing}
\newacronym{flops}{flops}{floating point operations}
\newacronym{CS}{CS}{compressed sensing}
\newacronym{MP}{MP}{message passing}
\newacronym{MPA}{MPA}{message passing algorithm}
\newacronym{DNN}{DNN}{deep neural network}
\newacronym{ASB}{ASB}{adaptively scaled belief}
\newacronym{LLR}{LLR}{log-likelihood ratio}
\newacronym{BAd-VAMP}{BAd-VAMP}{bilinear adaptive vector AMP}
\newacronym{LIP}{LIP}{linear inference problem}
\newacronym{AoA}{AoA}{angle of arrival}
\newacronym{LS}{LS}{least square}
\newacronym{mmWave}{mmWave}{millimeter-wave}
\newacronym{RRI}{RRI}{right-rotationally invariant}
\newacronym{PDE}{PDE}{partial differential equation}
\newacronym{SC-VAMP}{SC-VAMP}{score-based VAMP}
\newacronym{DSM}{DSM}{denoising score matching}
\newacronym{SISO}{SISO}{soft-input soft-output}
\newacronym{TAP}{TAP}{Thouless–Anderson–Palmer}
\newacronym{EC}{EC}{expectation-consistent}
\newacronym{LDPC}{LDPC}{low-density parity-check}
\newacronym{EXIT}{EXIT}{extrinsic information transfer}
\newacronym{SLM}{SLM}{standard linear model}
\newacronym{GLM}{GLM}{generalized linear model}
\newacronym{PnP}{PnP}{plug-and-play}
\newacronym{RED}{RED}{regularization-by-denoising}
\newacronym{MI}{MI}{mutual information}
\newacronym{AutoDiff}{AutoDiff}{automatic differentiation}
\newacronym{KL}{KL}{Kullback-Leibler}
\newacronym{BG}{BG}{Bernoulli-Gaussian}
\newacronym{MLP}{MLP}{multi-layer perceptron}
\newacronym{GPU}{GPU}{graphics processing unit}
\newacronym{VJP}{VJP}{vector-Jacobian product}
\newacronym{ULA}{ULA}{unadjusted Langevin algorithm}

\newtheorem{theorem}{Theorem}

\newtheorem{remark}{Remark}

\newcommand{\E}{\mathbb{E}}
\newcommand{\N}{\mathcal{N}}

\DeclareMathOperator{\Tr}{Tr}

\begin{document}

\title{Score-Based VAMP  \\with Fisher-Information-Based Onsager Correction}

\author{
\IEEEauthorblockN{
Tadashi Wadayama\IEEEauthorrefmark{1},
Takumi Takahashi\IEEEauthorrefmark{2}
}
\IEEEauthorblockA{\IEEEauthorrefmark{1}Nagoya Institute of Technology, \texttt{wadayama@nitech.ac.jp}}
\IEEEauthorblockA{\IEEEauthorrefmark{2}The University of Osaka, \texttt{takahashi@comm.eng.osaka-u.ac.jp}}
}

 \maketitle

\begin{abstract}
We propose score-based VAMP (SC-VAMP), a variant of vector approximate message passing (VAMP) in which the Onsager correction is expressed and computed via conditional Fisher information, thereby enabling a Jacobian-free implementation.
Using learned score functions, SC-VAMP constructs nonlinear \ac{MMSE} estimators through Tweedie’s formula and derives the corresponding Onsager terms from the score-norm statistics, avoiding the need for analytical derivatives of the
prior or likelihood.
When combined with random orthogonal/unitary mixing to mitigate non-ideal, structured or correlated sensing settings, the proposed framework extends VAMP to complex black-box inference problems where explicit modeling is intractable.
Finally, by leveraging the entropic \ac{CLT}, we provide an information-theoretic perspective on the Gaussian approximation underlying \ac{SE}, offering insight
into the decoupling principle beyond idealized \ac{i.i.d.} settings, including nonlinear regimes.
\end{abstract}

\begin{IEEEkeywords}
VAMP, inverse problem, score function, denoising score matching, fisher information, Onsager correction, decoupling 
\end{IEEEkeywords}

\glsresetall

\section{Introduction}
\label{sec:introduction}

High-dimensional inverse problems—estimating a signal vector $\bm{x} \in \mathbb{R}^N$ from noisy, possibly nonlinear measurements $\bm{y} \in \mathbb{R}^M$—are ubiquitous in modern science and engineering, spanning fields from wireless communications and compressed sensing to medical imaging and computational photography.
Traditionally, these problems have been approached using two distinct paradigms: model-based iterative algorithms and data-driven deep learning approaches.

Model-based algorithms, particularly \ac{AMP}~\cite{donoho2009}  and closely related approaches such \ac{VAMP}~\cite{rangan2019} and \ac{OAMP}~\cite{ma2017}, offer powerful and analytically tractable frameworks for signal recovery.
Under ideal conditions—such as \ac{i.i.d.} Gaussian sensing matrices or \ac{RRI} matrices—their asymptotic performance is rigorously characterized by a scalar recursion known as \ac{SE}, which accurately predicts the trajectory of the reconstruction error.
In this work, we deliberately focus on the Bayes-optimal formulation of \ac{VAMP}, which combines strong empirical performance with a broad range of practical applicability, \textit{i.e.}, \ac{RRI} sensing matrices.

However, Bayes-optimal \ac{VAMP} faces significant challenges in real-world scenarios. 
Deriving the optimal \ac{MMSE} denoiser and its associated divergence
(the Onsager correction term\footnote{Strictly speaking, the extrinsic information exchange mechanism in \ac{VAMP} differs from the Onsager correction originally derived for \ac{AMP}. \ac{AMP} explicitly subtracts a self-interference term arising from iterative feedback, whereas \ac{VAMP} enforce orthogonality between estimation errors across modules to achieve statistical decoupling. Since in both cases the decoupling correction can be expressed via the denoiser divergence, we refer to this divergence-based term as an Onsager correction throughout this paper.}) typically requires analytical knowledge of both the prior distribution $p_X$ and the likelihood $p_{Y|X}$.
In practice, the observation model itself is often highly structured or constrained—for example, through convolutional operators, array manifolds, or forward models based on \acp{PDE}~\cite{wadayama-2025-1}.
%
Moreover, in many modern applications, the prior is complex (\textit{e.g.}, natural images), and the front-end observation involves deterministic nonlinearities (\textit{e.g.}, sensor saturation or optical processes) in real-world systems.
Under such conditions, the analytical derivation of optimal estimators and the associated \ac{SE} becomes computationally intractable or may no longer be valid.

In contrast, data-driven approaches, such as \ac{DU}~\cite{hershey2014deep,balatsoukas2019deepunfolding,LISTA,TISTA}, replace the analytical components of iterative algorithms with trainable parameters or trainable neural networks. 
While empirically successful, these methods often operate as black boxes: they typically lack the theoretical guarantees enjoyed by their model-based counterparts, and it is often unclear whether performance bottlenecks arise from fundamental physical limits or from insufficient training.
Furthermore, training such end-to-end models frequently requires massive datasets of paired input-output samples $(\bm{x}, \bm{y})$, which may not always be available or easy to generate in practice, particularly when the forward model corresponds to a complex physical system.

In this paper, we bridge this gap by proposing {\em \ac{SC-VAMP}}, 
a framework that integrates the analytically tractable structure of \ac{VAMP} with the expressive power of score-based models~\cite{song2021}.
Our key insight is that the essential components of Bayes-optimal \ac{VAMP}—namely, the nonlinear \ac{MMSE} estimator and the Onsager correction term—can be completely characterized by the \textit{score function} $\nabla \log p(\cdot)$ and the \textit{Fisher information} $J=\mathbb{E}[\|\nabla \log p(\cdot)\|^2]$.
This abstraction enables a unified treatment of both linear and nonlinear problems under a single mathematical formalism, while also providing a conceptual bridge between statistical-physics-inspired analyses, Bayesian inference, and information-theoretic viewpoints in a coherent manner.

By leveraging \ac{DSM}~\cite{vincent2011}, we learn these quantities directly from data, thereby enabling the construction of approximately optimal estimators without requiring explicit analytical expressions for the underlying density functions.
Moreover, we show that, in the classical linear Gaussian setting, 
\ac{SC-VAMP} reduces to standard Bayes-optimal \ac{VAMP} and recovers the same \ac{SE} fixed point and mutual information.
As a result, \ac{SC-VAMP} remains fully consistent with existing theory, while extending its applicability far beyond linear Gaussian models.

The contributions of this paper are summarized as follows:

\begin{itemize}
  \item \textbf{Data-Driven Generalization of Bayes-optimal VAMP:}
  We formulate \ac{SC-VAMP}, in which the denoiser is implemented via Tweedie's formula using a learned score function, and the Onsager correction is computed from the estimated Fisher information.
  When combined with random orthogonal/unitary mixing (\textit{i.e.}, continuous-valued random mixing) to handle non-ideal sensing settings, \ac{SC-VAMP} extends \ac{VAMP} to arbitrary complex priors and nonlinear observation models while retaining the core message-passing structure, without requiring explicit analytical derivatives.

  \item \textbf{Jacobian-Free Onsager via Fisher Information (Score-Centric View of VAMP):}
  We show that the Onsager correction can be expressed solely in terms of the conditional Fisher information along the pseudo-\ac{AWGN} channel, \textit{e.g.}, $\alpha(v)=1-\frac{v}{N}J$, where the Fisher information $J$ is estimated from the squared norm of the conditional score.
  This eliminates the need for explicit Jacobian/divergence evaluations, which are typically required in \ac{AMP}/\ac{VAMP} when using complex or learned denoisers. 
  As a result, we obtain a unified score-centric interface in which both 
  the posterior mean update (via Tweedie's formula) and the Onsager term are determined by the same object—the conditional score.
  We further provide complementary rationales based on Stein's identity and on an information-theoretic route (via de~Bruijn and I--MMSE relationship), offering an interpretable view of the Onsager term as an information-consistency (curvature) correction.

  \item \textbf{Theoretical Rationale via Entropic CLT:}
  We provide an information-theoretic perspective on the Gaussian approximation underlying \ac{SE}. By leveraging the entropic \ac{CLT}~\cite{barron1986entropy} and the results by Tulino and Verd\'{u}~\cite{tulino2006}, we interpret the linear mixing step in \ac{SC-VAMP} as a ``Gaussianizer,'' which tends to reduce the non-Gaussianity of estimation errors in terms of \ac{KL}-divergence.
  This information-theoretic viewpoint, when combined with random orthogonal/unitary mixing, offers insight into the decoupling principle in nonlinear and structured settings.

\end{itemize}

\section{Preliminaries}
\label{sec:preliminaries}

\subsection{Notation}

Unless otherwise stated, all logarithms are natural (nats) in this paper.
Bold letters denote vectors and matrices.
A random variable is denoted by an uppercase letter. 
For a random variable $U$ with density $p_U$, we write $h(U)$ for 
its differential entropy, $I(\cdot;\cdot)$ 
for mutual information \cite{cover2006}. 
The Fisher information \cite{fisher1922}\cite{cover2006} is defined via the score function 
$s_U(\bm u)\equiv\nabla_{\bm u}\log p_U(\bm u)$ as
$
J_U\equiv \mathbb{E}_{\bm u\sim p_U(\cdot)}\!\left[\|\nabla_{\bm u}\log p_U(\bm u)\|^2\right].
$

\subsection{Problem Setup}

In this paper, we investigate a score-based representation for 
the \ac{SISO} modules within the \ac{VAMP} framework. 
The objective of our {\em \acf{SC-VAMP}} framework is to estimate an unknown signal $X (\in \mathbb{R}^N) \sim p_X$ from an observation $Y \in \mathbb{R}^M$, which follows a general conditional probability distribution $p_{Y|X}$.
This formulation provides a broad modeling framework that encompasses conventional estimation problems based on standard linear observation models, while extending the applicability to arbitrary nonlinear 
and non-Gaussian systems.

\subsection{Related Works}
\label{subsec:related_works}


The proposed \ac{SC-VAMP} framework is closely related to a broad line of research on iterative inference algorithms originating from the \ac{TAP} equations~\cite{Thouless1977} introduced in spin-glass theory, as well as to their asymptotic analyses.
The concept of the Onsager reaction term, originally introduced in the \ac{TAP} equations, has since been abstracted and understood as an essential correction that ensures self-consistency in large-scale iterative inference algorithms, thereby enabling an asymptotic characterization via \ac{SE}.
Representative algorithms that can be regarded as culminating points of this development are \ac{AMP} and \ac{VAMP}, together with \ac{SE} as the asymptotic analytical framework that is intrinsically paired with these algorithms.

\ac{AMP} and \ac{VAMP} can be formulated as top-down algorithmic constructions based on generalized error models~\cite{rangan2019,Bayati2011,Bayati2015,Takeuchi2020}.
The rigorous validity of \ac{SE} for these algorithms has been established under appropriate matrix ensembles and regularity conditions, building on arguments closely related to conditioning techniques developed by Erwin Bolthausen in the context of spin-glass theory.
At the same time, these algorithms admit an alternative interpretation as large-system limits of Bayesian \ac{BP} (\textit{i.e.}, the sum–product algorithm) on dense graphs~\cite{Kabashima2003}.
From this perspective, the Onsager reaction term can be interpreted as a correction that compensates for correlations induced between messages during iterative updates.
This structural interpretation is consistent with the classical result that the \ac{TAP} equations can be derived from \ac{BP} in fully connected Ising models, and it can also be explained from the viewpoint of \ac{EC} approximations~\cite{Opper2005} and approximate free-energy minimization~\cite{Minka2001}.

In particular, in Bayes-optimal settings where the prior distribution and the likelihood are matched to the true generative model, the self-consistent fixed-point equations obtained via the replica method~\cite{Tanaka2002,Guo2005} (within the regime where the replica-symmetric assumption is valid) coincide with the fixed points of \ac{SE}.
In this case, the \ac{SE} recursion can be interpreted as a successive-substitution procedure for solving these fixed-point equations.
When the fixed point is unique and \ac{SE} converges, the asymptotic solution reached by the algorithm coincides with the Bayes-optimal estimator in the large-system limit~\cite{Bayati2011,Bayati2015}.
In this sense, Bayes-optimal \ac{AMP} and Bayes-optimal \ac{VAMP} can be viewed as occupying a conceptual nexus at which top-down algorithmic design principles rooted in spin-glass theory intersect with bottom-up interpretations based on Bayesian approximate inference.


These developments are also closely connected to earlier notions of decoupling and scalar-equivalent channel representations in coding theory, such as density evolution for \ac{LDPC} codes~\cite{ru2008book} and the turbo principle~\cite{tenbrink2001}.
In particular, the turbo principle—originally introduced as an algorithmic heuristic with empirically favorable convergence behavior—relies on the exchange of extrinsic information between constituent modules and is commonly analyzed via scalar-equivalent channels and \ac{EXIT} charts~\cite{Hagenauer2004}.
From a modern perspective, it is precisely the statistical-mechanics-inspired framework that provides a rigorous theoretical foundation for such principles, enabling discussions of optimality and convergence in asymptotic regimes, including both large-system limits in inference problems and infinite-length limits in coding theory.

Together with rigorous asymptotic analyses based on \ac{SE},
\ac{AMP} and \ac{VAMP} have been extended to \acp{GLM}~\cite{rangan2011gamp,Schniter2016,Zou2018,Meng2018,Cademartori2024}, providing powerful and analytically tractable inference frameworks.
However, with only a few highly restricted exceptions, it remains difficult to derive closed-form algorithms for the complex prior distributions and observation models with deterministic nonlinearities that commonly arise in practical engineering problems.

In a related but distinct line of work, learning-based denoisers have been
integrated into \ac{AMP}/\ac{VAMP}-type architectures.
\Ac{PnP}~\cite{venkatakrishnan2013pnp} and \ac{RED}~\cite{romano2017red,metzler2016damp} replace model-based priors with powerful image denoisers, achieving strong
empirical performance in practice.
Furthermore, numerous deep-unfolded \ac{AMP}/\ac{VAMP} architectures have been proposed, in which denoisers are implemented as neural networks and trained in an end-to-end manner.
\Acf{DU}~\cite{hershey2014deep} and its applications to communication systems~\cite{balatsoukas2019deepunfolding} demonstrate that unfolding a small number of message-passing iterations
into a trainable network can yield remarkably strong empirical performance~\cite{LISTA,TISTA}.
However, these approaches often come at the cost of a clean theoretical interpretation grounded in \ac{SE}.
Since learned denoisers and their associated Jacobians (Onsager terms) are not explicitly tied to any underlying probabilistic model, it becomes difficult to quantitatively assess how far the algorithmic performance lies from information-theoretic limits, or whether \ac{SE} remains valid in such settings.

\subsection{Our Contributions}
\label{subsec:contribution}

Against this background, score-based generative models and \ac{DSM}~\cite{vincent2011,song2021} have recently emerged as a powerful frameworks for learning gradients of log-densities (\textit{i.e.}, score function) directly from data.
While several works have explored the use of learned score functions for denoising, sampling, and as building blocks for Bayesian inference, most of these approaches are not explicitly connected to the statistical estimation theory underlying \ac{AMP}/\ac{VAMP}-style algorithms, particularly to analyses based on \ac{SE}.

Our work can be viewed as {\em bringing these two lines of research together}: we show that both the nonlinear \ac{MMSE} estimator and the Onsager correction in \ac{VAMP} can be entirely parameterized by the score function and its associated Fisher information.
Moreover, while these quantities can be learned directly from data via \ac{DSM}, the resulting framework remains fully consistent with the scalar Gaussian optimality enjoyed by standard \ac{VAMP} in classical settings.
As a result, this work substantially extends the applicability of \ac{VAMP}, establishing a framework that reconciles learning-based nonlinear estimation with asymptotic analyses via \ac{SE}, in regimes where asymptotic decoupling is approximately satisfied.

In addition, the theoretical foundations of this work—most notably the integral representations of \ac{MI}~\cite{wadayama2025}—are deeply rooted in the interplay between information theory and thermodynamics.
These connections trace back to Stam's use of the heat equation~\cite{stam1959}, were further developed through the entropy \ac{CLT} due to Brown~\cite{brown1982} and Barron~\cite{barron1986entropy}, and were ultimately unified by the I–\ac{MMSE} relationship established by Guo et al.~\cite{guo2005}.
This framework provides a coherent conceptual link between statistical-physics-inspired analytical methods, Bayesian inference, and information-theoretic perspectives.


\section{Score-Based VAMP}

In this section, we present the theoretical framework underlying \ac{SC-VAMP}.
By reformulating Bayes-optimal \ac{VAMP} through the lens of score functions and Fisher information, we obtain a unified algorithmic framework in which optimal nonlinear estimators are constructed
via Tweedie’s formula.
This formulation extends the classical linear setting to general nonlinear regimes, while preserving a principled and analytically interpretable structure.

\subsection{Score-Based Representation of SISO Modules}

We begin by describing a \ac{SISO} module, which serves as a fundamental building block of the proposed \ac{SC-VAMP} presented below.
The \ac{SISO} module acts as an estimator: given an input vector $\bm x_{in} \in \mathbb{R}^N$, representing the prior information, an (optional) observation vector $\bm y \in \mathbb{R}^M$, and an input variance $v_{in} > 0$, , it produces an extrinsic output $\bm x_{out} \in \mathbb{R}^N$ and an extrinsic variance $v_{out} > 0$.

The score-based representation of a \ac{SISO} module used 
in the \ac{SC-VAMP} is given as follows:
\begin{align}
  \label{eq:tweedie}
\bm x_{post} &= \bm x_{in} + v_{in} s(\bm x_{in}| \bm y),\\
\label{eq:extrinsic_output}
\bm x_{out} &= \frac{\bm x_{post} - \alpha(v_{in}) \bm x_{in}}{1 - \alpha(v_{in})}, \\
\label{eq:se_eq}
v_{out} &= \frac{\alpha(v_{in})}{1 - \alpha(v_{in})} v_{in},   \\
\label{eq:onsager}
\alpha(v_{in}) &= 1 - \frac{v_{in}}{N} J_{X_{in} |Y},  
\end{align}
where $\bm x_{post}$ denotes the posterior mean, and $\alpha(v_{in})$ is the Onsager coefficient.
The (conditional) score function $s(\cdot | \cdot)$ is defined by
\begin{equation}
s(\bm x_{in}|\bm y) \equiv  \nabla_{\bm x_{in}} \log p_{X_{in}|Y}(\bm x_{in}| \bm y).
\end{equation}
The corresponding conditional Fisher information is
\begin{align}
    J_{X_{in}|Y} \equiv {\mathbb E}[\|s(\bm x_{in}| \bm y)\|^2],
\end{align}
where $Y$ and $X_{in}$ denote the random variables corresponding to the observation vector and the input to the \ac{SISO} module, respectively.
In addition, the posterior variance is given by
\begin{align}
v_{post}  = v_{in} \alpha(v_{in}).
\end{align} 

The expressions above describe a \ac{SISO} module that depends on $\bm y$.
For a \ac{SISO} module that does not use $\bm y$ (\textit{i.e.}, {\em denoiser module}, we use the unconditional score function: $s(\bm x_{in})\equiv  \nabla_{\bm x_{in}} \log p_{X_{in}}(\bm x_{in})$ is place of $s(\bm x_{in}| \bm y)$.
The derivation and further discussion of these relations are provided in the following subsections.

Beyond computational savings, expressing the Onsager correction through conditional Fisher information also suggests an information-geometric interpretation: the strength of the correction can be viewed as being governed by the local curvature of the induced posterior.

\subsection{Score-based VAMP (SC-VAMP) Algorithm}

The proposed \ac{SC-VAMP} algorithm operates by iteratively applying two \ac{SISO} modules, that exchange extrinsic information in the form of mean vectors and variances.
We refer to these modules as Module A and Module B, whose operations are specified below.

\textbf{Module A (Observation Module):} This module typically incorporates the observation $\bm y \in \mathbb{R}^M$ together with the channel conditional \ac{PDF} $p_{Y|X}$ or an associated likelihood function.
Given inputs $\bm x_{in, A} \in \mathbb{R}^N$ and $v_{in, A} \in \mathbb{R}$, it computes extrinsic outputs $\bm x_{out, A} \in \mathbb{R}^N$ and $v_{out, A} \in \mathbb{R}$ using the conditional score function $s(\bm x_{in} | \bm y)$:
\begin{align}
\label{eq:extrinsic}
\bm x_{out,A} &= \frac{v_{in,A} s(\bm x_{in,A}| \bm y)+ (1 - \alpha_A(v_{in,A}) )\bm x_{in,A}}{1 - \alpha_A(v_{in,A})}, \\
  v_{out,A}&= \frac{\alpha_A(v_{in,A})}{1 - \alpha_A(v_{in,A})} v_{in,A},  \\
  \alpha_A(v_{in,A}) &= 1 - \frac{v_{in,A}}{N} J_{X_{in,A} |Y},  
\end{align}
with message updates
\begin{align}
  \bm x_{in,A} &= \bm x_{out,B},\quad v_{in,A} = v_{out,B}.
\end{align}

\textbf{Module B (Denoiser Module):} This module typically incorporates prior information through the prior distribution $p_X$.
Given inputs $\bm x_{in, B}$ and $v_{in, B}$, it computes extrinsic outputs $\bm x_{out, B}$ and $v_{out, B}$ using the score function $s(\bm x_{in})$, which does not depend on $\bm y$:
\begin{align}
    \bm x_{out,B} &= \frac{v_{in,B} s(\bm x_{in,B})+ (1 - \alpha_B(v_{in,B}) )\bm x_{in,B}}{1 - \alpha_B(v_{in,B})}, \\
  v_{out,B} &= \frac{\alpha_B(v_{in,B})}{1 - \alpha_B(v_{in,B})} v_{in,B},  \\
  \alpha_B(v_{in,B}) &= 1 - \frac{v_{in,B}}{N} J_{X_{in,B}},   
\end{align}
with message updates
\begin{align}
  \bm x_{in,B} &= \bm x_{out,A},\quad v_{in,B} = v_{out,A}.
\end{align}

The two modules exchange extrinsic messages iteratively for a fixed number of iterations $T$ or until the extrinsic parameters converge.
The algorithm is typically initialized by setting the output of Module B at iteration $t=0$ to initial values.
Specifically, we set $\bm x_{out, B}^{(0)} = \bm 0$ and $v_{out, B}^{(0)} = v_{init}$, where $v_{init}$ denotes the variance of the prior distribution or, more generally, a large initial variance for numerical stability.

\subsection{Derivation of SISO Module}

\subsubsection{Problem Setting}

In this subsection, we provide the derivation and justification for the score-based \ac{SISO} module equations introduced earlier.
The \ac{SC-VAMP} algorithm consists of two \ac{SISO} modules; however, due to the inherent symmetry of the algorithm, 
the discussion in this section focuses on a single generic \ac{SISO} module.

We consider the problem of estimating an unknown signal $X \sim p_X$, which is the target of the estimation.
We assume two types of soft information regarding $X$ are available.


\textbf{Soft-Input 1 (Prior Information):} An observation of $X \in \mathbb{R}^N$ corrupted by additive Gaussian noise $Z \sim {\cal N}(\bm 0, v_{in} \bm I_N)$, \textit{i.e.},
\begin{equation}
    X_{in} = X + Z.
\end{equation}
This can be interpreted as observing $X$, drawn from the prior distribution $p_X$, through a {\em virtual} additive Gaussian noise channel. 
%
An realization of $X_{in}$ is denoted by $\bm x_{in}$, which serves as the input to the \ac{SISO} module.

\textbf{Soft-Input 2 (Observation Information):} An observation $Y \in \mathbb{R}^M$ follows:
\begin{equation}
    X \sim p_X,\quad Y \sim p_{Y|X}.
\end{equation}
A representative example is a nonlinear measurement model in which the signal $X$ is passed through a nonlinear mapping $f:\mathbb{R}^N \to \mathbb{R}^M$ and corrupted by additive Gaussian noise $W \sim {\cal N}(\bm 0, \sigma_N^2 \bm I_M)$, \textit{i.e.},
\begin{equation}
    Y = f(X) + W.
\end{equation}
%
where we assume that $X$, $Z$, and $W$ are mutually independent.

In the following, we first derive the posterior mean (\ac{MMSE} estimate) of $X$, \textit{i.e.}, $\mathbb{E}[X |X_{in} = \bm x_{in}, Y = \bm y]$, based on these two inputs $X_{in}$ and $Y$.

\subsubsection{Conditional Tweedie's Formula}

The core component of the \ac{SISO} module is the \ac{MMSE} estimator given by the posterior expectation $\mathbb{E}[X |X_{in} = \bm x_{in}, Y = \bm y]$.
This estimator admits a concise representation in terms of the score function, which can be viewed as a conditional generalization of Tweedie’s formula~\cite{robbins1956,efron2011} to the case where side information $Y$ is available.

The posterior expectation is derived as follows:
\begin{align} 
    \mathbb{E}[X |\bm x_{in}, \bm y] 
    &= \int \bm x p(\bm x | \bm x_{in}, \bm y) d\bm x \\
    &= \frac{1}{p(\bm x_{in}, \bm y)} \int \bm x p(\bm x_{in}, \bm y | \bm x) p(\bm x) d\bm x \\
    &= \frac{1}{p(\bm x_{in}, \bm y)} \int \bm x p(\bm x_{in} | \bm x) p(\bm y | \bm x) p(\bm x) d\bm x,
\end{align}
where the last equality follows from the conditional independence $X_{in} \perp Y$ given $X$.
Here, we focus on the observation model $p(\bm x_{in} | \bm x)$, whose \ac{PDF} is given by
\begin{equation}
    p(\bm x_{in} | \bm x) = \frac{1}{(2\pi v_{in})^{N/2}} 
    \exp \left( - \frac{\| \bm x_{in} - \bm x \|^2}{2 v_{in}} \right).
\end{equation}
Taking the partial derivative with respect to $\bm x_{in}$, we obtain
\begin{align*}
    v_{in} \nabla_{\bm x_{in}} p(\bm x_{in} | \bm x) &= - (\bm x_{in} - \bm x) p(\bm x_{in} | \bm x),
\end{align*}
which can be rearranged as
\begin{equation}
    \bm x p(\bm x_{in} | \bm x) = \bm x_{in} p(\bm x_{in} | \bm x) + v_{in} \nabla_{\bm x_{in}} p(\bm x_{in} | \bm x). \label{eq:stein_identity}
\end{equation}
Substituting this identity into the numerator of the posterior expectation integral yields
\begin{align*}
    &\int \bm x p(\bm x_{in} | \bm x) p(\bm y | \bm x) p(\bm x) d\bm x \\
    &= \int \left( \bm x_{in} p(\bm x_{in} | \bm x) + v_{in} \nabla_{\bm x_{in}} p(\bm x_{in} | \bm x) \right) p(\bm y | \bm x) p(\bm x) d\bm x \\
    &= \bm x_{in} \underbrace{\int p(\bm x_{in} | \bm x) p(\bm y | \bm x) p(\bm x) d\bm x}_{p(\bm x_{in}, \bm y)} \\
    &+ v_{in} \int \nabla_{\bm x_{in}} p(\bm x_{in} | \bm x) p(\bm y | \bm x) p(\bm x) d\bm x.
\end{align*}
In the second term, exchanging integration and differentiation (under standard regularity conditions) gives
\begin{align*}
    &\int \nabla_{\bm x_{in}} p(\bm x_{in} | \bm x) p(\bm y | \bm x) p(\bm x) d\bm x \\
    &= \nabla_{\bm x_{in}} \int p(\bm x_{in} | \bm x) p(\bm y | \bm x) p(\bm x) d\bm x \\
    &= \nabla_{\bm x_{in}} p(\bm x_{in}, \bm y).
\end{align*}
Combining these results, we obtain
\begin{align*}
    \mathbb{E}[X | \bm x_{in}, \bm y] &= \frac{1}{p(\bm x_{in}, \bm y)} \left( \bm x_{in} p(\bm x_{in}, \bm y) + v_{in} \nabla_{\bm x_{in}} p(\bm x_{in}, \bm y) \right) \\
    &= \bm x_{in} + v_{in} \frac{\nabla_{\bm x_{in}} p(\bm x_{in}, \bm y)}{p(\bm x_{in}, \bm y)} \\
    &= \bm x_{in} + v_{in} \nabla_{\bm x_{in}} \log p(\bm x_{in}, \bm y)\\
    &= \bm x_{in} + v_{in} \nabla_{\bm x_{in}} (\log p(\bm x_{in}| \bm y)+\log p(\bm y))\\
    &= \bm x_{in} + v_{in} s(\bm x_{in}|\bm y),
\end{align*}
where the last equality follows from $\nabla_{\bm x_{in}}\log p(\bm y) = 0$.
This completes the derivation of the conditional Tweedie's formula, corresponding to \eqref{eq:tweedie}; therefore, in the following, we denote the posterior mean by $\bm x_{post} = \mathbb{E}[X | \bm x_{in}, \bm y]$, which serves as the \ac{MMSE} estimate of $X$ given the soft inputs $X_{in}$ and $Y$.

\subsubsection{Derivation of Extrinsic Output with Onsager Correction}

The information passed from a \ac{SISO} module to the subsequent stage is required to be \textit{extrinsic}, meaning that the contribution of the input information (prior) should be appropriately removed.
In this subsection, we derive \eqref{eq:extrinsic_output}, which realizes this extrinsic message construction, based on a Gaussian approximation and its reproducibility properties.

Specifically, we approximate the posterior estimate (total information) produced by the \ac{SISO} module by a Gaussian distribution ${\cal N}(\bm x_{post}, v_{post} \bm I_N)$ with mean $\bm x_{post}$ and variance $v_{post}$.
Under this approximation, the posterior distribution can be expressed as the product (equivalently, the sum of precisions) of the Gaussian input distribution ${\cal N}(\bm x_{in}, v_{in} \bm I_N)$ and an extrinsic Gaussian distribution ${\cal N}(\bm x_{out}, v_{out} \bm I_N)$ generated within the module~\cite{Minka2001}.
This leads to the relations:
\begin{align}
    \frac{1}{v_{post}} &= \frac{1}{v_{in}} + \frac{1}{v_{out}} \label{eq:precision_sum}, \\
    \frac{\bm x_{post}}{v_{post}} &= \frac{\bm x_{in}}{v_{in}} + \frac{\bm x_{out}}{v_{out}}. \label{eq:mean_sum}
\end{align}
Here, the posterior estimation error variance $v_{post}$ can be interpreted as the per-symbol \ac{MMSE} in regimes where a scalar-equivalent channel approximation is accurate.
Introducing the Onsager coefficient $\alpha$, defined via the average divergence of the estimator (as discussed later), we write
\begin{equation}
    v_{post} = v_{in} \alpha(v_{in}) \label{eq:mmse_alpha},
\end{equation}
where $\alpha(v_{in}) < 1$ reflects the effective variance reduction induced by the denoising operation.

\subsubsection{Derivation of Extrinsic Variance $v_{out}$}

Substituting \eqref{eq:mmse_alpha} into \eqref{eq:precision_sum} and rearranging terms, the extrinsic variance $v_{out}$ is obtained by solving
\begin{align}
\label{eq:extrinsic_variance}
    \frac{1}{v_{out}} &= \frac{1}{v_{in} \alpha(v_{in})} - \frac{1}{v_{in}} 
    = \frac{1 - \alpha(v_{in})}{v_{in} \alpha(v_{in})} 
\end{align}
This yields \eqref{eq:se_eq}.
The resulting expression serves as the variance update equation for the extrinsic message and is consistent with the variance recursion that appears in \ac{SE} analyses under appropriate decoupling assumptions.

\subsubsection{Derivation of Extrinsic Mean \texorpdfstring{$\bm x_{out}$}{x_out}}

Similarly, multiplying both sides of \eqref{eq:mean_sum} by $v_{post} = v_{in}\alpha(v_{in})$ gives:
\begin{equation}
    \bm x_{post} = \alpha(v_{in}) \bm x_{in} + \frac{v_{in}\alpha(v_{in})}{v_{out}} \bm x_{out}
\end{equation}
Using \eqref{eq:extrinsic_variance}, the coefficient simplifies to
\begin{equation}
    \frac{v_{in}\alpha(v_{in})}{v_{out}} = v_{in}\alpha(v_{in}) \cdot \frac{1-\alpha(v_{in})}{v_{in}\alpha(v_{in})} = 1 - \alpha(v_{in}),
\end{equation}
which leads to
\begin{equation}
    \bm x_{post} = \alpha(v_{in}) \bm x_{in} + (1 - \alpha(v_{in})) \bm x_{out}.
\end{equation}
Solving this expression for $\bm x_{out}$ yields the extrinsic output equation with Onsager correction in \eqref{eq:extrinsic_output}.
This operation corresponds to the Onsager correction, which subtracts the input contribution ($\alpha \bm x_{in}$) 
from the posterior mean and rescales the residual accordingly.

\subsubsection{Derivation of the Onsager Coefficient $\alpha$ (Divergence Route)}

We first recall a standard derivation of the Onsager coefficient $\alpha$ that is commonly used in the literature.
In the \ac{VAMP} framework, $\alpha$ is defined as the average divergence of the \ac{MMSE} estimator $\eta(\bm x_{in}) \equiv \bm x_{in} + v_{in} s(\bm x_{in}|\bm y)$, namely,
\begin{equation}
\label{eq:score_Onsager}
    \alpha(v_{in}) \equiv \frac{1}{N} \E [ \nabla_{\bm x_{in}} \cdot \eta(\bm x_{in}) ] 
    = 1 + \frac{v_{in}}{N} \E [ \nabla_{\bm x_{in}} \cdot s(\bm x_{in}|\bm y) ].
\end{equation}
Applying Stein’s identity~\cite{stein1981}, $\E[\nabla \cdot s] = - \E[\|s\|^2]$, 
the divergence term can be rewritten in terms of the squared norm of the score function (\textit{i.e.}, conditional Fisher information), yielding:
\begin{align}
    \alpha(v_{in}) &= 1 - \frac{v_{in}}{N} \E [ \| s(\bm x_{in}|\bm y) \|^2 ] \\
    &= 1 - \frac{v_{in}}{N} J_{X_{in} |Y},
\end{align}
which leads to \eqref{eq:onsager}. It is well-known that we can easily derive the identity \eqref{eq:mmse_alpha} 
directly from the definition of the Onsager coefficient \eqref{eq:score_Onsager}.

\subsubsection{Derivation of Onsager Coefficient $\alpha$ (Information-Theoretic Route)}

It is also possible to show the identity on the Onsager coefficient \eqref{eq:mmse_alpha} via another {\em information-theoretic route}.
From the conditional de Bruijn's identity~\cite{stam1959}, the derivative of the conditional \ac{MI} satisfies
\begin{align}
\label{eq:de_Bruijins}
    \frac{d}{dv_{in}} I(X; X_{in}|Y) &= \frac{1}{2}J_{X_{in}|Y} - \frac{N}{2v_{in}}.
\end{align}
On the other hand, the same derivative can be expressed via the I–MMSE relationship~\cite{guo2005} as
\begin{align}
\label{eq:immse}
  \frac{d}{dv_{in}} I(X; X_{in}|Y) 
  &= -\frac{1}{2v_{in}^2}\mathrm{mmse}(v_{in}), 
\end{align}
where $\mathrm{mmse}(v_{in})\equiv \mathbb{E}[\|X - \mathbb{E}[X | X_{in}, Y]\|^2]$ denotes the \ac{MMSE} of estimating $X$ from $X_{in}$ and $Y$.
Equating \eqref{eq:de_Bruijins} and \eqref{eq:immse} yields
\begin{align}
  \mathrm{mmse}(v_{in}) = Nv_{in} - v_{in}^2 J_{X_{in}|Y}.
\end{align}
Recalling that the posterior variance $v_{post}$ is the per-symbol \ac{MMSE}, we obtain
\begin{align}
    v_{post} = \frac{1}{N}\mathrm{mmse}(v_{in}) 
    &= v_{in} - \frac{v_{in}^2}{N} J_{X_{in}|Y}\\
    &= v_{in}\left(1 - \frac{v_{in}}{N} J_{X_{in}|Y} \right) \\
    &= v_{in} \alpha(v_{in}). \label{eq:Onsager_correction}
\end{align}
If the Onsager coefficient $\alpha(v_{in})$ is defined as the correction factor in the relation \eqref{eq:Onsager_correction}, the same expression for $\alpha(v_{in})$ can be obtained without explicitly evaluating the divergence.
Alternatively, $\alpha(v_{in})$ can be interpreted as a constraint required to satisfy the information-theoretic consistency between the \ac{MMSE} and Fisher information.

\subsection{Recovery of Bayes-optimal VAMP for Linear Observations}
\label{sec:linear_vamp_recovery}

We confirm that the score-based formulation proposed in this paper recovers the standard Bayes-optimal \ac{VAMP}~\cite{rangan2019} in the special case of linear Gaussian observations, \textit{i.e.}, \acp{SLM}.

Consider the observation model given by a \ac{SLM}:
\begin{equation}
    \bm y = \bm A \bm x + \bm n, \quad \bm n \sim \mathcal{N}(\bm 0, \sigma_w^2 \bm I).
\end{equation}
The input to the \ac{SISO} module is $\bm x_{in} = \bm x + \bm z, \bm z \sim \mathcal{N}(\bm 0, v_{in} \bm I_N)$.
In this setting, the random variables $(\bm x, \bm y, \bm x_{in})$ follow a joint Gaussian distribution, and the corresponding conditional score function is therefore linear.

In this system, the posterior mean of $p(\bm x | \bm x_{in}, \bm y)$, \textit{i.e.}, the \ac{MMSE} estimate, is given by the well-known Wiener filter, \textit{i.e.}, \ac{LMMSE} estimator:
\begin{equation}
    \hat{\bm x}_{LMMSE} = \left( \frac{1}{v_{in}} \bm I + \frac{1}{\sigma_w^2} \bm A^\top \bm A \right)^{-1} \left( \frac{1}{v_{in}} \bm x_{in} + \frac{1}{\sigma_w^2} \bm A^\top \bm y \right).
    \label{eq:lmmse_def}
\end{equation}
On the other hand, working backward from Tweedie's formula $\eta(\bm x_{in}) = \bm x_{in} + v_{in} s(\bm x_{in} | \bm y)$, the score function in this case must satisfy
\begin{equation}
    s(\bm x_{in} | \bm y) = \frac{\hat{\bm x}_{LMMSE} - \bm x_{in}}{v_{in}}.
\end{equation}
This expression is linear in $\bm x_{in}$.

In standard \ac{VAMP}, for a linear estimator of the form $\eta(\bm x_{in}) = \bm K \bm x_{in} + \bm c$, the Onsager coefficient is calculated using the trace of the Jacobian as
\begin{equation}
    \alpha_{VAMP} = \frac{1}{N} \Tr\left( \frac{\partial \eta}{\partial \bm x_{in}} \right) = \frac{1}{N} \Tr(\bm K).
\end{equation}
From \eqref{eq:lmmse_def}, the gain matrix $\bm K$ in this case, using the posterior covariance matrix $\bm \Sigma_{post} = (v_{in}^{-1} \bm I + \sigma_w^{-2} \bm A^\top \bm A)^{-1}$, is given by
\begin{equation}
    \bm K = \frac{1}{v_{in}} \bm \Sigma_{post}.
\end{equation}
Accordingly, the Onsager coefficient used in standard \ac{VAMP} can be obtained by
\begin{equation}
    \alpha_{VAMP} = \frac{1}{N v_{in}} \Tr(\bm \Sigma_{post}).
    \label{eq:alpha_standard}
\end{equation}

Next, we check the score-based definition of the Onsager coefficient in \eqref{eq:score_Onsager}, and have the following identity:
\begin{align}
    \alpha_{score} &= 1 + \frac{v_{in}}{N} \E [ \nabla_{\bm x_{in}} \cdot s(\bm x_{in} | \bm y) ] \notag \\
    &= 1 + \frac{v_{in}}{N} \E \left[ \nabla_{\bm x_{in}} \cdot \left( \frac{\eta(\bm x_{in}) - \bm x_{in}}{v_{in}} \right) \right] \notag \\
    &= 1 + \frac{v_{in}}{N} \left( \frac{1}{v_{in}} \E [ \nabla \cdot \eta ] - \frac{1}{v_{in}} \E [ \nabla \cdot \bm x_{in} ] \right) \notag \\
    &= 1 + \frac{1}{N} \E [ \Tr(\bm K) ] - \frac{1}{N} N \notag \\
    &= 1 + \alpha_{VAMP} - 1 \notag \\
    &= \alpha_{VAMP}.
\end{align}
This calculation shows that, in the linear Gaussian setting, the Onsager coefficient obtained from the score-based formulation coincides exactly with that used in standard \ac{VAMP}.
Therefore, the formulation of the \ac{SC-VAMP} is a complete generalization of standard Bayes-optimal \ac{VAMP}.

\subsection{Recommended Implementation: Efficient Estimation of Fisher Information}
\label{subsec:implementation}

A significant computational advantage of the proposed \ac{SC-VAMP} framework over standard \ac{VAMP} lies in the estimation of the Onsager correction term $\alpha(v_{in})$.
In standard \ac{VAMP}, calculating $\alpha(v_{in})$ requires evaluating the divergence of the denoiser $\eta(\cdot)$, defined as $\frac{1}{N}\sum_{n=1}^N \frac{\partial \eta_n}{\partial x_{in,n}}$.
For deep neural network-based denoisers, computing this divergence typically necessitates \ac{AutoDiff} to obtain the Jacobian trace, which is computationally expensive and memory-intensive, especially for high-dimensional signals.

In contrast, \ac{SC-VAMP} circumvents the need for explicit derivative computations by leveraging Stein's identity, which relates the divergence to the Fisher information.
Consequently, the Onsager coefficient can be determined solely from the squared norm of the score function output.
We propose the following efficient Monte Carlo estimation procedure using mini-batch processing, which is recommended for practical implementations.

Let $\mathcal{B} \equiv \{ (\mathbf{x}_{in}^{(i)}, \mathbf{y}^{(i)}) \}_{i=1}^{B}$ be a mini-batch of $B$ input samples at a given iteration, where $\mathbf{x}_{in}^{(i)}$ denotes the input to the \ac{SISO} module (prior or likelihood) and $\mathbf{y}^{(i)}$ denotes the observation (if applicable).
The Fisher information is estimated empirically as the average squared $\ell_2$-norm of the learned score function outputs:
\begin{equation}
\label{eq:lscore}
    \hat{J}_{\theta} = \frac{1}{B} \sum_{i=1}^{B} \left\| \mathbf{s}_{\theta}(\mathbf{x}_{in}^{(i)} | \mathbf{y}^{(i)}) \right\|^2,
\end{equation}
where $\mathbf{s}_{\theta}(\cdot)$ represents the learned score network.
Using this estimate, the estimate of Onsager coefficient is computed as
\begin{equation}
    \hat{\alpha}(v_{in}) = 1 - \frac{v_{in}}{N} \hat{J}_{\theta}.
\end{equation}
This approach reduces the computational complexity from the order of Jacobian computations to simple forward propagation followed by a norm calculation.

\subsection{Utilization of Pre-Trained Denoisers}

In many practical scenarios, a high-performance denoiser $\eta_{\text{opt}}(\cdot)$ (\textit{e.g.}, a pre-trained deep neural network like DnCNN or DRUNet) may already be available.
In such cases, it is possible to bypass the explicit training of a score model by deriving the \textit{implicit score} directly from the denoiser outputs.
Based on Tweedie's formula \eqref{eq:tweedie}, the score function corresponding to the denoiser $\eta_{\text{opt}}$ can be retrieved as
\begin{equation}
    \hat{\mathbf{s}}(\mathbf{x}_{in}) = \frac{\eta_{\text{opt}}(\mathbf{x}_{in}) - \mathbf{x}_{in}}{v_{in}}.
    \label{eq:reverse_tweedie}
\end{equation}
Assuming $\eta_{\text{opt}}$ closely approximates the \ac{MMSE} estimator, this implicit score $\hat{\mathbf{s}}$ is a consistent estimator of the true score $\nabla \log p(\mathbf{x}_{in})$.
Consequently, Stein's identity holds, and the Fisher information required for the Onsager correction can be estimated by substituting \eqref{eq:reverse_tweedie} into the Monte Carlo estimator:
\begin{equation}
    \hat{J}_{\eta} = \frac{1}{B} \sum_{i=1}^{B} \left\| \frac{\eta_{\text{opt}}(\mathbf{x}_{in}^{(i)}) - \mathbf{x}_{in}^{(i)}}{v_{in}} \right\|^2.
\end{equation}
This hybrid approach allows \ac{SC-VAMP} to leverage existing \ac{SotA} denoisers while retaining the efficient, derivative-free computation of the Onsager term.
Note that numerical stability requires care when $v_{in}$ becomes extremely small, although in typical VAMP operation, the convergence behavior usually prevents singularity issues.

\section{Optimality in Scalar Gaussian Channels}

It is well established from the decoupling principle of standard \ac{VAMP} that, in linear inverse problems, the iterative process can be characterized by an equivalent scalar Gaussian channel (\textit{i.e.}, an \ac{AWGN} channel) at each iteration.
Under the assumption of decoupling and Gaussian inputs, it therefore
suffices to examine whether the proposed score-based update rules recover the \ac{MI} $I(X;Y)$ in this scalar Gaussian setting. 
Accordingly, this section focuses on the analysis of the basic scalar \ac{AWGN} case.

\subsection{Optimality Theorem}

In this subsection, we explicitly derive the \ac{SE} fixed point by applying the \ac{SC-VAMP} update rules \eqref{eq:tweedie}-\eqref{eq:onsager} to a scalar Gaussian channel.
The following theorem states that the resulting \ac{SE} fixed point achieves the \ac{MI} of the underlying channel.

\begin{theorem}[Optimality in Scalar Gaussian Channels]
\label{thm:consistency}
Consider a scalar Gaussian channel $Y = X + Z$, where $X \sim \mathcal{N}(0, P)$ and noise $Z \sim \mathcal{N}(0, \sigma^2)$.
The \ac{MI} associated with the fixed point of \ac{SC-VAMP} satisfies
\begin{equation}
    I_{\mathsf{VAMP}} = I(X;Y) = \frac{1}{2} \log \left( 1 + \frac{P}{\sigma^2} \right),
\end{equation}
where $I_{\mathsf{VAMP}}$ denotes the \ac{MI} corresponding to the \ac{SE} fixed point.
\end{theorem}

\subsection{Proof}
We analyze the updates of Module A (Likelihood) and Module B (Prior) in sequence.

\subsubsection{Module A (Likelihood)}
The input to Module A is given by $X_{in,A} = X + W_A$, where $W_A \sim \mathcal{N}(0, v_{in,A})$.
For the Gaussian likelihood $p(y|x_{in})$, the conditional score function is linear and given by
\begin{equation}
\label{eq:linear_score}
    s_A(x_{in}|y) = \frac{y - x_{in}}{\sigma^2 + v_{in,A}}.
\end{equation}
The corresponding Fisher information $J_A$ with respect to $x_{in}$ is constant:
\begin{equation}
    J_A = \frac{1}{\sigma^2 + v_{in,A}}.
\end{equation}
Using the definition of the Onsager coefficient in Eq. \eqref{eq:onsager}, we obtain:
\begin{align}
    \alpha_A(v_{in,A}) 
    &= 1 - v_{in,A} J_A  \\
    &= 1 - \frac{v_{in,A}}{\sigma^2 + v_{in,A}} = \frac{\sigma^2}{\sigma^2 + v_{in,A}}.
    \label{eq:linear_div}
\end{align}
Substituting $\alpha_A$ into the variance update rule \eqref{eq:se_eq}, the extrinsic variance becomes
\begin{equation}
    v_{out,A} = v_{in,A} \frac{\alpha_A}{1 - \alpha_A} = v_{in,A} \frac{\frac{\sigma^2}{\sigma^2 + v_{in,A}}}{\frac{v_{in,A}}{\sigma^2 + v_{in,A}}} = \sigma^2.
\end{equation}
Thus, Module A outputs an extrinsic variance equal to the noise variance $\sigma^2$, independent of the input variance.

\subsubsection{Module B (Prior)}
The input to Module B is given by $X_{in,B} = X + W_B$, where $W_B \sim \mathcal{N}(0, v_{in,B})$.
For the Gaussian prior $X \sim \mathcal{N}(0, P)$, the score function is given by
\begin{equation}
    s_B(x_{in}) = - \frac{x_{in}}{P + v_{in,B}}.
\end{equation}
Similarly, the corresponding Fisher information is $J_B = \frac{1}{P + v_{in,B}}$.
Using \eqref{eq:onsager}, the Onsager coefficient becomes
\begin{equation}
    \alpha_B(v_{in,B}) = 1 - v_{in,B} J_B = \frac{P}{P + v_{in,B}}.
\end{equation}
Applying the variance update rule \eqref{eq:se_eq} yields
\begin{equation}
    v_{out,B} = v_{in,B} \frac{\alpha_B}{1 - \alpha_B} = P.
\end{equation}
Thus, Module B outputs an extrinsic variance equal to the signal power $P$.

\subsubsection{Fixed Point and Mutual Information}
The \ac{SE} recursion $v_{in,B}^{(t)} = v_{out,A}^{(t-1)}$ immediately converges to the fixed point $v_{in,B}^* = \sigma^2$.
At this fixed point, the effective channel seen by Module B is $X_{in,B} = X + Z_{eff}$, where $Z_{eff} \sim \mathcal{N}(0, \sigma^2)$.
The resulting \ac{MI} is given by
\begin{align}
    I_{\mathsf{VAMP}} &\equiv I(X; X_{in,B}) = \frac{1}{2} \log \left( 1 + \frac{P}{v_{in,B}^*} \right) \\
    &= \frac{1}{2} \log \left( 1 + \frac{P}{\sigma^2} \right)=I(X;Y).
\end{align}
This analytically confirms that the \ac{MI} associated with the \ac{SE} fixed point coincides with the true \ac{MI} $I(X;Y)$.

\subsubsection{Consistency of the Estimator}
While Theorem \ref{thm:consistency} establishes consistency at the level of 
the second moment, the \ac{SC-VAMP} also correctly recovers the optimal point estimator (first moment). 
To verify this, we examine the extrinsic mean messages at the fixed point. 
For Module A, from \eqref{eq:linear_score} and \eqref{eq:linear_div}, the extrinsic update rule \eqref{eq:extrinsic} yields
\begin{equation}
    x_{out,A} = \frac{
    v_{in,A}\left(y - x_{in}\right)+ v_{in,A}x_{in,A}
    }{
    v_{in,A} 
    }
    =y.
\end{equation}
Thus, the extrinsic message from the likelihood module reduces to the observation itself.
For Module B, which employs a zero-mean prior, the extrinsic message is $x_{out,B} = 0$.
Consequently, at the fixed point, Module B receives $x_{in,B} = y$ with effective noise variance $v_{in,B}^* = \sigma^2$.
The final estimate produced by the algorithm is therefore
\begin{equation}
    \hat{x}_{\mathsf{VAMP}} = \mathbb{E}[X \mid X_{in,B} = y] = \frac{P}{P + \sigma^2} y.
\end{equation}
This coincides with the \ac{LMMSE} estimator (Wiener filter) for the scalar Gaussian channel, indicating consistency in both information-theoretic limits and point estimation accuracy.

\begin{remark}[Generalization to Vector Linear Gaussian Models]
\rm 
The result established in Theorem \ref{thm:consistency} for the scalar case naturally extends to vector linear Gaussian models $\bm y = \bm A \bm x + \bm z$ with Gaussian priors.
%
A fundamental property of \ac{VAMP} is that, for an \ac{RRI} matrix $\bm{A}$ in the large-system limit, the estimation problem asymptotically decouples into a collection of parallel, independent scalar Gaussian channels.
Since the proposed score-based update rules recover the optimal behavior for each underlying scalar Gaussian channel, the \ac{SC-VAMP} algorithm is theoretically consistent with asymptotically achieving the corresponding \ac{MI} in the vector setting.
In particular, the resulting \ac{MI} is given by $I(\bm X; \bm Y) = \frac{1}{2} \sum_i \log ( 1 + \frac{\lambda_i P}{\sigma^2} ),$ where $\lambda_i$ denotes the eigenvalues of $\bm A^\top \bm A$.
\end{remark}

\section{Data-Driven Approach for SC-VAMP}
\label{subsec:dsm}

When the analytical form of the score function is available, both the denoiser and the Onsager coefficient can be derived in closed form.
For instance, in linear observation models, the score function admits an explicit analytical expression.
A key advantage of \ac{SC-VAMP} is that, even when such
analytical forms are unavailable, both the denoiser and the
Onsager coefficient can instead be obtained through a learned
score model.

\subsection{Denoising Score Matching (DSM)}

This becomes particularly powerful when the score function $s(\bm x_{in}) = \nabla_{\bm x_{in}} \log p(\bm x_{in})$ (for Module B) 
or $s(\bm x_{in} | \bm y) = \nabla_{\bm x_{in}} \log p(\bm x_{in} | \bm y)$ (for Module A) is not analytically tractable, which is common
in problems involving complex signal priors $p_X$ or nonlinear observation models $p_{Y|X}$.
In such settings, a \textit{data-driven} approach to inverse problems becomes feasible through learned score models.

In these cases, the true score function can be approximated by a parameterized model, typically a deep neural network $\bm s_{\theta}(\cdot)$, trained using \textit{\acf{DSM}}.
The core idea of \ac{DSM}~\cite{hyvarinen2005, vincent2011} is to learn parameters $\theta$ by minimizing the expected squared difference between the network output and the true score.
Although the true score $\nabla_{\bm x_{in}} \log p(\bm x_{in})$ is unknown, \ac{DSM} yields a tractable training objective.
For the denoiser module (Module B) with input noise variance $v_{in}$, the \ac{DSM} objective is
\begin{align}
    &J_{DSM}(\theta) \nonumber \\
    &= \E_{\bm x \sim p_X, \bm z \sim \mathcal{N}(0, v_{in} \bm I_N)} \left[ \left\| \bm s_{\theta}(\bm x + \bm z) - \nabla_{\bm x_{in}} \log p(\bm x_{in} | \bm x) \right\|^2 \right]
\end{align}
where $\bm x_{in} = \bm x + \bm z$.
The conditional score $\nabla_{\bm x_{in}} \log p(\bm x_{in} | \bm x)$ corresponds to the Gaussian noise model $\mathcal{N}(\bm x_{in} | \bm x, v_{in} \bm I_N)$, and is given analytically by
\begin{equation}
    \nabla_{\bm x_{in}} \log p(\bm x_{in} | \bm x) = -\frac{\bm x_{in} - \bm x}{v_{in}} = -\frac{\bm z}{v_{in}}.
\end{equation}
As a result, the objective simplifies to the computationally tractable loss:
\begin{equation}
    \mathcal{L}(\theta) = \E_{\bm x \sim P_X, \bm z \sim \mathcal{N}(0, v_{in} \bm I_N)} \left[ \left\| \bm s_{\theta}(\bm x + \bm z) + \frac{\bm z}{v_{in}} \right\|^2 \right],
\end{equation}
which requires only samples from the prior $P_X$.
A similar objective can be derived for learning the conditional score $\bm s_{\theta}(\bm x_{in} | \bm y)$.
It should be noted that a learned score function with small $v_{in}$ is unstable 
if the prior of $\bm x$ is not differentiable, \textit{e.g.}, a mixture of the Dirac's delta functions. For such a case,
Gaussian smoothing for $\bm x$ could mitigate the problem.

\subsection{Estimating Fisher Information}

Once the network $\bm s_{\theta}$ is trained, it serves as a
\textit{plug-in} score model within the \ac{SC-VAMP} algorithm.
The posterior mean is then computed via Tweedie's formula as
\begin{equation}
    \bm x_{post} = \bm x_{in} + v_{in} \bm s_{\theta}(\bm x_{in}|\bm y).
\end{equation}
The Onsager coefficient $\alpha(v_{in})$ is estimated using the Fisher information associated with the \textit{learned} score model.
Given the learned score model $s_{\bm \theta}(\cdot)$, the Fisher information can be approximated by Monte Carlo estimation:
\begin{align}
    J_{\theta} &= 
    \frac{1}{B}\sum_{i=1}^B \|s_{\bm \theta}(\bm{x}_{in,i}| \bm y_i)\|^2 \\
    &\simeq \E[\|\bm s_{\theta}(X_{in}|Y)\|^2], \\
    \alpha_{\theta}(v_{in})&= 1 - \frac{v_{in}}{N} J_{\theta},
\end{align}
where $\bm{x}_{in,i} \sim p_{X_{in}}$ and $\bm y_i \sim p_{Y}$ denote samples drawn from the corresponding distributions. 
In practice, $J_{\theta}$ can be estimated empirically during the \ac{VAMP} iterations by calculating the average squared norm of the network's output vector. 
This data-driven procedure enables the proposed \ac{SC-VAMP} framework to be applied to a much broader class of complex and nonlinear inference problems.

\section{Decoupling Principle and State Evolution}
\label{sec:decoupling}

The empirical success of iterative algorithms such as \ac{VAMP}
and AMP is underpinned by the \textit{decoupling principle}.
In the large-system limit ($M = \delta N \to \infty$ with a fixed ratio $\delta = M/N$), this principle states that the original $N$-dimensional vector problem statistically decomposes into $N$ parallel, independent scalar Gaussian channels.
This decomposition is fundamental, as it enables the iterative dynamics of the algorithm to be tracked by a simple one-dimensional recursion known as \ac{SE}.
In this section, we discuss the structural conditions that facilitate this decoupling—specifically, statistical symmetry and mechanisms that suppress short-range correlations, such as locally tree-like factor graphs or high-dimensional random mixing—and provide an information-theoretic justification for the Gaussian
approximation underlying \ac{SE}.

\subsection{Decoupling for SC-VAMP}

For \ac{SC-VAMP}, the decoupling principle implies that the input
to the nonlinear \ac{SISO} module (denoiser) $\eta(\cdot)$ at iteration $t$ can be modeled as $\bm{x}_{in}^{(t)} = \bm{x} + \bm{e}^{(t)}$, where $\bm{x}$ is the true signal and $\bm{e}^{(t)}$ is an error vector that is asymptotically \ac{i.i.d.} Gaussian, $\bm{e}^{(t)} \sim \N(\bm 0, v^{(t)} \bm{I}_N)$.

This decoupling allows the algorithm’s performance to be tracked by \ac{SE}, which reduces the high-dimensional dynamics to the scalar recursion of the variance $v^{(t)}$.
The \ac{SC-VAMP} framework—defined through Tweedie’s formula and the Fisher
information–based Onsager term—naturally adopts \ac{SE} as a central theoretical tool.
However, such decoupling is not universally guaranteed.
In general, the validity of the decoupling principle relies on two
key ingredients: (i) a form of statistical symmetry across coordinates, and (ii) a mechanism that suppresses short-range correlations, such as locally tree-like factor graphs or high-dimensional random mixing (\textit{e.g.}, \ac{RRI} linear transforms).

\subsection{Statistical Symmetry}

The first requirement, often referred to as \textit{identically distributed}, is a direct consequence of statistical symmetry, or permutation invariance.
If the system—such as the prior $p_X$, the observation operator $\bm{A}$, and the
nonlinear function $f$—is statistically invariant under permutations of its coordinates, then no coordinate $i$ is distinguished from any other coordinate $j$.
Consequently, the marginal distributions of the error components $e_i^{(t)}$ and $e_j^{(t)}$ must be identical:
\begin{equation}
    P(e_i^{(t)}) = P(e_j^{(t)}) \quad \forall i, j.
\end{equation}
This symmetry implies that tracking a single scalar variance $v(t) = {\mathbb{E}}[e_i^2]$ is sufficient, provided that the error components are also independent.

\subsection{Suppressing Short-Range Correlations}

The second ingredient is the suppression of short-range statistical dependencies between messages.
In sparse message-passing systems, this requirement is realized when 
the underlying \ac{FG} is {\em locally tree-like}~\cite{ru2008book}.
In iterative message-passing algorithm, correlations between messages arises from 
{\em short cycles (loops)} in the graph. If a message sent by a node can return 
to itself via a short path (\textit{e.g.}, $A \to B \to C \to A$), then the messages received by node $A$ from $B$ and $C$ will be correlated, as they share a recent common ancestor.
Such correlations violate the independence assumption underlying the decoupling principle.

A graph is said to be \textit{locally tree-like} if, in the limit $N \to \infty$, the probability of the existence of any short cycle vanishes.
In this case, messages arriving at a node are asymptotically independent.
Consequently, in sparse graphical models, local tree-likeness serves as the canonical mechanism for suppressing short-range correlations and ensuring effective decoupling.
In dense linear models, such as standard \ac{VAMP}, an analogous effect can be interpreted, from a Bayesian inference perspective, as being achieved through
high-dimensional random mixing, as discussed in the next subsection.

\subsection{Achieving Decoupling in Practice}

There exist practical systems that satisfy the two conditions.
Representative examples include the following.

\subsubsection{Component-wise Functions (\textit{e.g.}, \acp{GLM})}

When the nonlinearity acts component-wise, \textit{i.e.}, $y_i = f(x_i)$, the
associated \ac{FG} is inherently symmetric and perfectly sparse, with no cycles.
This represents the simplest setting in which the decoupling principle holds.

\subsubsection{Right-Rotationally Invariant Matrices (\textit{e.g.}, standard VAMP)}

For dense matrices used in \ac{VAMP}, the key property is \textit{\acf{RRI}}, meaning that the matrix’s right singular vectors are uniformly (Haar) distributed over the orthogonal group.
Unlike standard \ac{AMP} which relies on the \textit{i.i.d.} nature of matrix entries, \ac{VAMP} leverages this rotational symmetry.
As a result, the estimation error vector remains isotropic (statistically symmetric) across iterations.
Under standard large-system assumptions for \ac{RRI} matrices and Lipschitz denoisers, while it is not immediate that the joint distribution of all error coordinates is \textit{i.i.d.}, any finite-dimensional subset of coordinates becomes asymptotically \acl{i.i.d.}.
This weaker, marginal form of decoupling is sufficient to support the use of scalar \ac{SE} in the sense of an empirical fixed-point analysis, without explicitly relying on a locally tree-like graph structure.


\subsubsection{Random Interleaver (\textit{e.g.}, Turbo/LDPC Codes)}

This provides a key mechanism for handling structured systems.
Consider a function ${f}$ that represents a locally coupled system, in which each output component depends only on a small subset of input variables (\textit{i.e.}, the Jacobian of ${f}$ is sparse).
A typical example is convolution, which is sparse but highly structured and often lacks both statistical symmetry and a locally tree-like \ac{FG} due to fixed local patterns.
By \textit{sandwiching} ${f}$ with random permutations (interleavers), \textit{i.e.}, ${f}'(\bm{x}) = {\pi}^{-1} {f}({\pi} \bm{x})$, one can simultaneously:
\begin{itemize}
    \item {Restore symmetry:} The resulting system becomes statistically permutation-invariant.
    \item {Enforce tree-likeness:} The interleaver disrupts fixed local cycles in ${f}$, yielding a random sparse graph that, with high probability, becomes locally tree-like as $N \to \infty$.
\end{itemize}
This principle underlies the validity of density evolution for Turbo and \ac{LDPC} codes.

\subsubsection{Random Modulation (random orthogonal/unitary matrices)}

This approach can be viewed as a generalization of random interleaving to linear inverse problems.
As discussed in recent studies~\cite{Liu2025}, for an arbitrary channel matrix $\bm A$—which may be deterministic and structured—right-multiplication by a random unitary matrix $\bm \Xi$ (random modulation) yields an equivalent
channel $\bm J = \bm A \bm \Xi$.
The transformed matrix $\bm J$ shares the same singular values as $\bm A$ but has asymptotically Haar-distributed singular vectors.
As a result, the estimation error becomes statistically isotropic.
From an inference perspective, this random modulation can be interpreted as placing $\bm J$ in the same \textit{universality class} as \ac{RRI} matrices, suggesting that \ac{SE} predictions remain accurate even for structured channels.
Conceptually, this operation acts as a generalized linear interleaver that \textit{whitens} spatial correlations in the channel.

\subsection{Information-Theoretic Perspective via Entropic CLT}
\label{subsec:entropic_clt}

\Ac{SE} tracks the evolution of second-order statistics, such as the variance $v(t)$, under the implicit assumption that the effective estimation error can be approximated by a Gaussian distribution.
Importantly, the validity of \ac{SE} does not require convergence of the full joint distribution of the error vector to a Gaussian law; rather, it relies on a weaker form of convergence, namely the empirical convergence of finite-dimensional marginals, which is sufficient to characterize pseudo-Lipschitz observables via a law-of-large-numbers effect.
From this perspective, the entropic \ac{CLT} provides a complementary, information-theoretic viewpoint on the Gaussian approximation commonly invoked in \ac{SE} analyses. 
%
Whereas the classical \ac{CLT} guarantees convergence in distribution, the entropic \ac{CLT} establishes a stronger notion of convergence in terms of information measures. 
Specifically, Barron~\cite{barron1986entropy} showed that for a sum of independent random variables, the \ac{KL}-divergence to the corresponding Gaussian distribution decreases monotonically as the number of summands increases.
Tulino and Verd\'{u}~\cite{tulino2006} further analyzed this phenomenon using the I-MMSE relationship,, quantifying the reduction of non-Gaussianity from an information-theoretic standpoint.

In \ac{SC-VAMP}, the linear mixing step—implemented via multiplication by an orthogonal matrix $\bm{V}^T$ or a random modulator $\bm{\Xi}$—forms weighted sums
of the $N$ input error components.
Under the decoupling assumption, whereby these inputs are asymptotically independent, the entropic \ac{CLT} provides an information-theoretic perspective suggesting that the distribution of each output component becomes increasingly Gaussian in the \ac{KL}-divergence sense as the system size $N$ grows.
From this viewpoint, the linear mixing step can be interpreted as acting as \textit{Gaussianizer}, attenuating structured non-Gaussian effects introduced by the nonlinear denoiser 
Accordingly, for each fixed coordinate $i$ and any fixed iteration index $t$, the marginal distribution of the effective noise may approach a Gaussian distribution in the sense that
\begin{equation}
    D_{\text{KL}}(P_{e_i^{(t)}} \| \mathcal{N}(0, v^{(t)})) \to 0 \quad \text{as } N \to \infty,
\end{equation}
under appropriate independence and mixing assumptions.
While such entropic convergence is not required for \ac{SE} to hold, it offers additional intuition for the Gaussian approximation commonly employed in \ac{SE} and for the use of second-order moments as effective summary statistics in empirical fixed-point analysis.

More generally, one may insert random orthogonal/unitary transforms and their inverses around both modules A and B, so that each nonlinear block operates in a randomly mixed coordinate system while the overall end-to-end mapping remains unchanged.
Such double-sided random mixing has the potential to homogenize and decorrelate the estimation errors seen by both modules, thereby facilitating decoupling, even when both the ``channel'' and ``prior'' modules involve complex nonlinearities.
A rigorous characterization of these schemes is left for future work.

Along a related direction, \textit{\ac{SE}-aware batch normalization} and \textit{deep-unfolded Onsager corrections} may allow \ac{SE} to serve as a useful design guideline even beyond the classical large-system setting.
In particular, \ac{SE}-based theory may offer systematic insights into improving performance in nonlinear and finite-dimensional systems with correlated sources.

\section{Numerical Experiments}

In this section, we present numerical experiments to validate the proposed \ac{SC-VAMP} framework.

\subsection{Linear Observation System}

In this subsection, we present numerical experiments to validate the proposed \ac{SC-VAMP} framework for a linear observation model.
Importantly, in these experiments, we use \emph{analytical score functions} derived from the known prior and likelihood distributions, rather than learned score functions obtained via \ac{DSM}.
This allows us to isolate the validation of the algorithmic framework itself from potential errors introduced by score function approximation.

\subsubsection{Experimental Setup}

We consider a linear observation model $\bm{y} = \bm{A}\bm{x} + \bm{n}$, where $\bm{x} \in \mathbb{R}^N$ is the unknown signal, $\bm{A} \in \mathbb{R}^{M \times N}$ is a \ac{RRI} sensing matrix, and $\bm{n} \sim \mathcal{N}(\bm{0}, \sigma_n^2 \bm{I}_M)$ is \ac{AWGN}.
The signal follows a \ac{BG} prior:
\begin{equation}
  p_X(x) = (1-\rho)\delta(x) + \rho \cdot \mathcal{N}(x; 0, \sigma_x^2),
\end{equation}
where $\rho$ denotes the sparsity rate.
The \ac{RRI} matrix is constructed as $\bm{A} = \bm{U}\bm{D}\bm{V}^\top$, where $\bm{U} \in \mathbb{R}^{M \times M}$ and $\bm{V} \in \mathbb{R}^{N \times N}$ are orthogonal matrices obtained via QR decomposition of random Gaussian matrices, and $\bm{D}$ is a diagonal matrix with singular values set to unity.

The experimental parameters are the following: signal dimension $N = 2000$, measurement dimension $M = 1000$ (compression ratio $\delta = 0.5$), sparsity rate $\rho = 0.1$, signal variance $\sigma_x^2 = 1.0$, and observation SNR $= 20$ dB.

\subsubsection{Mini-batch Implementation}

We employ a mini-batch implementation with batch size $B = 200$ to estimate the Fisher information required for the Onsager correction.
Specifically, the Fisher information $J$ is estimated as the sample average of the squared score norm over the mini-batch:
$
  \hat{J} = \frac{1}{B} \sum_{i=1}^{B} \|\bm{s}(\bm{x}_{\mathrm{in}}^{(i)} | \bm{y}^{(i)})\|^2,
$
where $\bm{s}(\cdot|\cdot)$ denotes the score function.
This mini-batch estimation corresponds to \eqref{eq:lscore} in the proposed framework and enables efficient computation of the Onsager coefficient $\alpha$ without explicit Jacobian calculations.

\subsubsection{Results}

\begin{figure}[htbp]
  \centering
  \includegraphics[width=1.0\columnwidth]{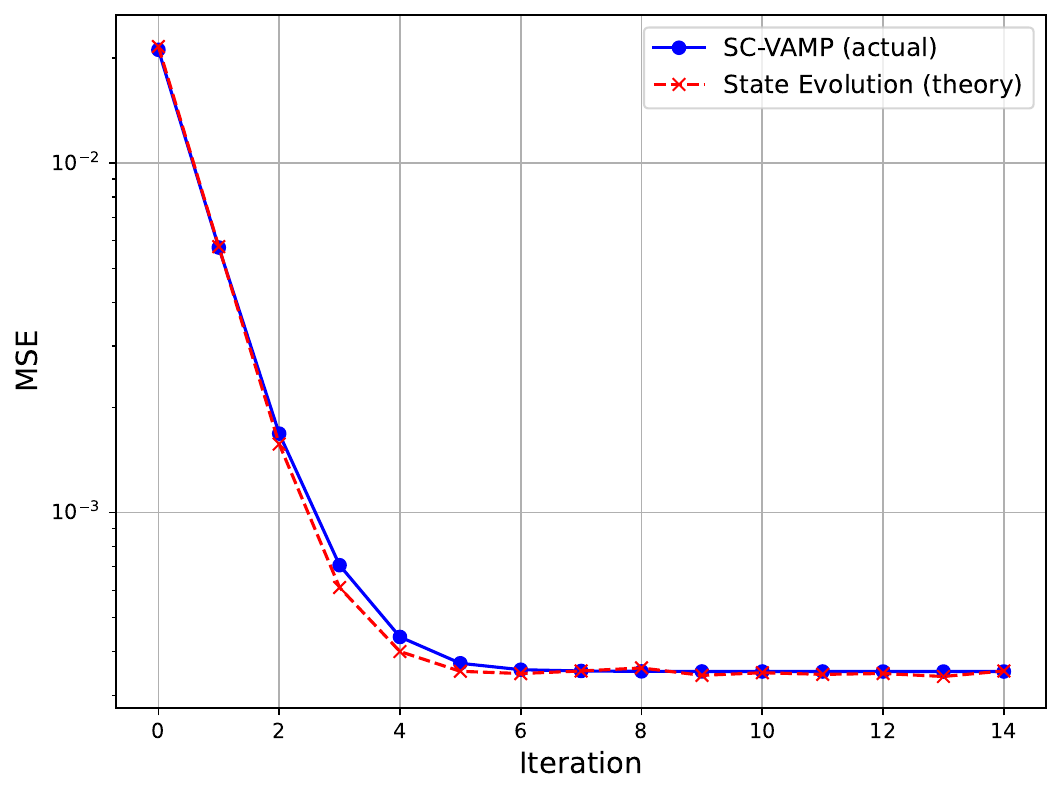}
  \caption{MSE convergence of SC-VAMP (actual) versus SE (theory) for a linear observation system with Bernoulli-Gaussian prior ($N=2000$, $M=1000$, $\rho=0.1$, SNR$=20$\,dB, batch size $B=200$).}
  \label{fig:mse_convergence}
\end{figure}

Figure~\ref{fig:mse_convergence} shows the \ac{MSE} convergence of \ac{SC-VAMP} compared with the theoretical prediction from \ac{SE}.
The actual \ac{MSE} trajectory (blue solid line) closely matches the \ac{SE} prediction (red dashed line), demonstrating that the proposed score-based formulation with mini-batch Fisher information estimation accurately reproduces 
the theoretically predicted behavior.

\begin{figure}[t]
  \centering
  \includegraphics[width=1.0\columnwidth]{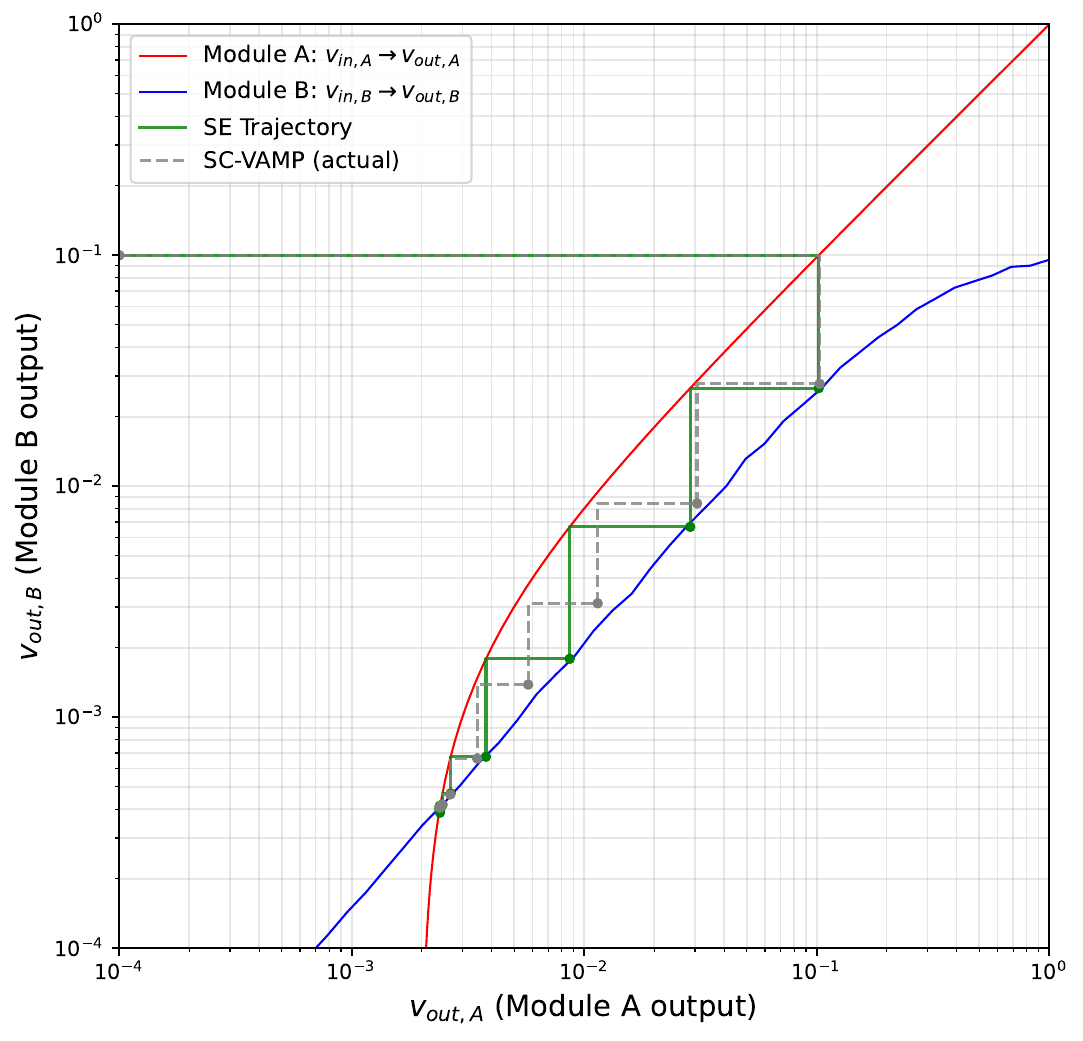}
  \caption{EXIT-style analysis of SC-VAMP showing Module~A and Module~B transfer characteristics, along with the SE trajectory (green) and actual SC-VAMP trajectory (gray dashed).}
  \label{fig:exit_analysis}
  \vspace{-2mm}
\end{figure}

Figure~\ref{fig:exit_analysis} presents an \ac{EXIT}-style analysis of \ac{SC-VAMP}.
The red curve represents the transfer characteristic of Module~A (observation module), mapping input variance $v_{\mathrm{in},A}$ to output variance $v_{\mathrm{out},A}$.
The blue curve represents the transfer characteristic of Module~B (denoiser module with \ac{BG} prior).
The green solid line shows the \ac{SE} trajectory, while the gray dashed line shows the actual \ac{SC-VAMP} trajectory.
Both trajectories follow the characteristic curves and converge to the intersection point, which corresponds to the fixed point of the \ac{SE} recursion.

These results confirm that (i) the score-based \ac{SISO} module formulation correctly implements the \ac{MMSE} estimator via Tweedie's formula, (ii) the mini-batch estimation of Fisher information provides an accurate Onsager correction, and (iii) the \ac{SC-VAMP} algorithm converges to the theoretically predicted fixed point.

\subsection{Correlated Prior with Learned Score Function}

In this subsection, we demonstrate the effectiveness of \ac{SC-VAMP} with a \emph{learned} score function for a correlated prior distribution.
Unlike the previous experiment using analytical scores, this experiment validates the complete data-driven pipeline including \ac{DSM} training and mini-batch Stein calibration.

\subsubsection{Correlated 2D Gaussian Prior}

We consider a structured prior where the signal $\bm{x} \in \mathbb{R}^N$ is partitioned into $N/2$ pairs of correlated Gaussian variables.
Each pair $(x_{2k-1}, x_{2k})$ for $k = 1, \ldots, N/2$ follows a bivariate Gaussian distribution:
\begin{equation}
  \begin{pmatrix} x_{2k-1} \\ x_{2k} \end{pmatrix} \sim \mathcal{N}\left(\bm{0}, \begin{pmatrix} \sigma^2 & \xi\sigma^2 \\ \xi \sigma^2 & \sigma^2 \end{pmatrix}\right),
\end{equation}
where $\sigma^2$ is the marginal variance and $\xi \in (-1, 1)$ is the correlation coefficient.
This prior captures pairwise dependencies that cannot be efficiently exploited by element-wise denoisers, making it a suitable testbed for evaluating learned score functions that respect the correlation structure.

\subsubsection{Pairwise Score Network Architecture}

To capture the pairwise correlation structure, we employ a neural network that processes pairs of variables jointly.
The score network $s_\theta: \mathbb{R}^2 \times \mathbb{R}_+ \to \mathbb{R}^2$ takes as input a pair $(r_1, r_2)$ along with the noise level $\sigma$, and outputs the corresponding score vector $(s_1, s_2)$.

The network architecture consists of a \ac{MLP} with the following structure:
\begin{itemize}
  \item Input layer: $3$ neurons (pair values $r_1, r_2$ and noise level $\sigma$)
  \item Hidden layers: $3$ fully-connected layers with $128$ neurons each
  \item Activation function: Softplus
  \item Output layer: $2$ neurons (score values $s_1, s_2$)
\end{itemize}

\subsubsection{Denoising Score Matching Training}

The score network is trained via \ac{DSM}.
Given samples $\bm{x}_0$ from the prior and noise $\bm{z} \sim \mathcal{N}(\bm{0}, \sigma^2 \bm{I})$ with $\sigma$ uniformly sampled from $[\sigma_{\min}, \sigma_{\max}] = [0.01, 3.0]$, the \ac{DSM} loss is:
\begin{equation}
  \mathcal{L}_{\mathrm{DSM}}(\theta) = \mathbb{E}_{\bm{x}_0, \sigma, \bm{z}}\left[\left\| s_\theta(\bm{x}_0 + \bm{z}, \sigma) + \frac{\bm{z}}{\sigma^2} \right\|^2\right].
\end{equation}
Training is performed using the Adam optimizer with learning rate $10^{-3}$ and cosine annealing schedule over $20,000$ iterations with batch size $256$.

\subsubsection{Mini-batch Stein Calibration}

A key component of our implementation is the mini-batch Stein calibration \cite{wadayama2025}, which enforces a necessary Stein moment constraint via a scalar rescaling.
For each \ac{SISO} module evaluation, we compute a scalar calibration coefficient $c$ using the Stein identity:
\begin{equation}
  c = \frac{-N}{(1/B)\sum_{i=1}^{B} \bm{r}^{(i)\top} s_\theta(\bm{r}^{(i)}, \sigma)},
\end{equation}
where $\{\bm{r}^{(i)}\}_{i=1}^B$ is the current mini-batch of inputs.
This calibration ensures that $\mathbb{E}[\bm{r}^\top (c \cdot s_\theta(\bm{r}, \sigma))] = -N$, which is a necessary condition for the score function of an $N$-dimensional Gaussian-corrupted distribution.
The calibrated denoiser output is then computed as:
\begin{equation}
  \bm{x}_{\mathrm{post}} = \bm{r} + v_{\mathrm{in}} \cdot c \cdot s_\theta(\bm{r}, \sqrt{v_{\mathrm{in}}}),
\end{equation}
where $v_{\mathrm{in}}$ is the input variance. This mini-batch calibration is applied consistently in both the \ac{SC-VAMP} algorithm and the \ac{SE} computation, ensuring fair comparison.

\subsubsection{Experimental Setup}

The experimental parameters are: signal dimension $N = 2000$, measurement dimension $M = 1000$ (compression ratio $\delta = 0.5$), prior standard deviation $\sigma = 1.0$, correlation coefficient $\xi = 0.9$, observation SNR $= 20$\,dB, and mini-batch size $B = 1000$. The sensing matrix $\bm{A}$ is constructed 
as an \ac{RRI} matrix identical to the previous experiment.

\subsubsection{Results}

\begin{figure}[htbp]
  \centering
  \includegraphics[width=1.0\columnwidth]{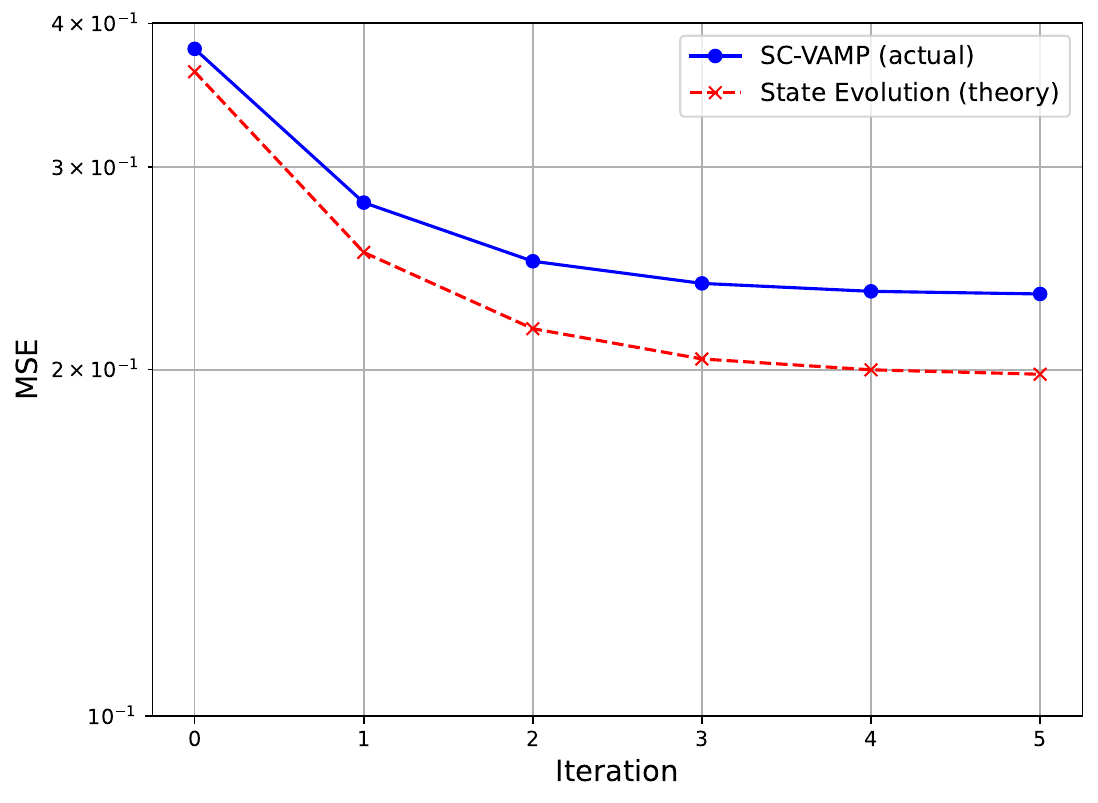}
  \caption{MSE convergence of SC-VAMP with learned pairwise score function for correlated 
  2D Gaussian prior ($N=2000$, $M=1000$, $\xi=0.9$, SNR$=20$\,dB). 
  The learned score network captures the pairwise correlation structure.}
  \label{fig:correlated_gaussian}
\end{figure}

Figure~\ref{fig:correlated_gaussian} shows the \ac{MSE} convergence for the correlated 2D Gaussian prior with $\xi = 0.9$.
Both \ac{SC-VAMP} (blue circles) and \ac{SE} (red crosses) exhibit monotonic decrease and converge within $6$ iterations.
The final \ac{MSE} values are $0.233$ (\ac{SC-VAMP}) and $0.198$ (\ac{SE}), with a relative difference of approximately $17$\%.

The gap between \ac{SC-VAMP} and \ac{SE} can be attributed to: (i) imperfect score function learning, and 
(ii) finite sample effects.
Nevertheless, both curves show consistent convergence behavior, demonstrating that the learned score function successfully captures the correlation structure and enables effective signal recovery.

This experiment validates that the \ac{SC-VAMP} framework with learned score functions can handle non-trivial correlation structures in the prior distribution---a scenario where traditional element-wise denoisers would fail to exploit the available statistical dependencies.

\section{Conclusion}
\label{sec:conclusion}

In this paper, we have proposed the \ac{SC-VAMP} framework, a data-driven approach to high-dimensional inverse problems.
By integrating message-passing algorithms with learned score-based models, we have
established a framework that broadens the applicability of standard \ac{VAMP} to settings that go beyond the classical linear Gaussian case.

A fundamental advantage of the proposed method is its universality with respect to the underlying probability distributions.
Unlike standard \ac{VAMP}, which often relies on linear Gaussian assumptions or requires analytical derivation of denoisers and their Jacobians, \ac{SC-VAMP} leverages learned score functions to construct (near) optimal nonlinear
estimators in a data-driven manner.
This enables applications to a broader class of problems, including:
\begin{itemize}
  \item \textbf{Arbitrary priors:} Correlated, complex, data-driven priors, such as natural images or semantic latent representations, for which analytical modeling is challenging.
  \item \textbf{General nonlinear observations:} Systems governed by deterministic nonlinear maps $\bm y = f(\bm x)$, such as sensor saturation, optical nonlinearities, or black-box physical processes.
\end{itemize}
Through this abstraction, \ac{SC-VAMP} can be viewed as a natural extension of \ac{VAMP} from specialized linear Gaussian settings toward more general inference scenarios encountered in practice.

Our analysis also provides a score-based reinterpretation of standard \ac{VAMP}, 
unifying the estimator update and the Onsager correction through conditional score and Fisher quantities, which offers a principled interface for extending
\ac{VAMP} to learned modules.
While \ac{SC-VAMP} is formulated using score functions in this paper, extending the framework to velocity-field framework (\textit{e.g.}, flow matching~\cite{lipman2023flowmatching} and rectified flow~\cite{liu2023rectifiedflow}) is a promising future direction.
Since there exists a one-to-one algebraic correspondence between the score function and the velocity field under a given probability path, an \ac{SC-VAMP}–like algorithm may be constructed using a velocity-field network to implement the \ac{MMSE} estimator.
In practice, velocity-field training is often observed to be more stable than score-based training, particularly in low-noise regimes.
Deriving an \ac{SC-VAMP} formulation within the rectified flow paradigm would be 
an interesting future problem, as it could bridge message-passing algorithms with the principles of $L_2$ optimal transport.

\section*{Acknowledgments}
This work was supported by JST, CRONOS, 
Japan Grant Number JPMJCS25N5 and
JSPS KAKENHI Grant Number JP25H01111.
The authors would like to thank Tomoharu Furudoi, M.S. student
at The University of Osaka, for his constructive comments on
an early draft of this manuscript.

\appendix
\section*{Hybrid Approach with Automatic Differentiation}
\label{sec:autodiff}

While the data-driven approach using \ac{DSM} described in Section \ref{subsec:dsm} is powerful for unknown or complex priors, many engineering applications involve a well-defined, deterministic observation model. 
In such cases, treating the observation process as a black box to be learned might be inefficient.
In this Appendix section, we propose a \textit{hybrid} architecture that combines the learned score function for the prior (Module B) with a computation-based \ac{MMSE} estimator with \ac{AutoDiff}~\cite{Baydin2018} for the observation model (Module A).

\subsection{Differentiable Observation Models}

Consider the nonlinear observation model:
\begin{equation}
    \bm{y} = f(\bm{x}) + \bm{n}, \quad \bm{n} \sim \mathcal{N}(\bm{0}, \sigma_n^2 \bm{I}_M),
\end{equation}
where $f: \mathbb{R}^N \to \mathbb{R}^M$ is a known, differentiable function representing the physical measurement process (\textit{e.g.}, channel nonlinearity, partial differential equation solver~\cite{wadayama-2025-1}, 
or rendering equation).
Unlike the prior distribution $p_{X}$, the conditional log likelihood $p(\bm{y}|\bm{x})$ is explicitly known:
\begin{equation}
    \log p(\bm{y}|\bm{x}) = -\frac{1}{2\sigma_n^2} \|\bm{y} - f(\bm{x})\|^2 + \text{const}.
\end{equation}
Modern deep learning frameworks (\textit{e.g.}, PyTorch, TensorFlow, JAX) allow 
for the efficient computation of the gradient of this likelihood with respect to $\bm{x}$ using \ac{AutoDiff}, specifically via the \ac{VJP}, without explicitly instantiating the Jacobian matrix:
\begin{equation} \label{eq:vjp}
    \nabla_{\bm{x}} \log p(\bm{y}|\bm{x}) = \frac{1}{\sigma_n^2} [\bm{J}_f(\bm{x})]^\top (\bm{y} - f(\bm{x})),
\end{equation}
where $J_f(\bm{x})$ denotes the Jacobian matrix of $f(\bm{x})$ at $\bm{x}$.

\subsection{Posterior Sampling for MMSE Estimation}

For Module A, the goal is to compute the \ac{MMSE} estimate (or equivalently, the conditional score $s_A(\bm{x}_{in}|\bm{y})$) given the input $\bm{x}_{in} = \bm{x} + \bm{z}$, where $\bm{z} \sim \mathcal{N}(\bm{0}, v_{in}\bm{I})$.
The \ac{MMSE} estimator is given by the posterior expectation:
\begin{equation}
    \bm{x}_{post} = \mathbb{E}[\bm{X} | \bm{x}_{in}, \bm{y}] = \int \bm{x} \frac{p(\bm{y}|\bm{x}) \mathcal{N}(\bm{x}_{in}|\bm{x}, v_{in}\bm{I})}{p(\bm{y}, \bm{x}_{in})} d\bm{x}.
\end{equation}
Since the integral involving the nonlinear $f(\bm{x})$ is generally intractable, 
we can approximate this expectation using Monte Carlo sampling driven by Langevin dynamics.
The gradient of the log-posterior required for Langevin sampling is given by
\begin{equation}
    \nabla_{\bm x} \log p(\bm x|\bm x_{\text{in}}, \bm y) 
    = \nabla_{\bm x} \log p(\bm y|\bm x)
      - \frac{\bm x - \bm x_{\text{in}}}{v_{\text{in}}}.
\end{equation}
Specifically, we employ the \ac{ULA} to generate samples.
The discrete-time update rule at iteration $k$ with step size $\delta$ is given by
\begin{align} \nonumber
    \bm{x}^{(k+1)} &= \bm{x}^{(k)} + \delta 
    \left( \nabla_{\bm{x}} \log p(\bm{y}|\bm{x}^{(k)}) - \frac{\bm{x}^{(k)} - \bm{x}_{in}}{v_{in}} \right) \\
     &+ \sqrt{2\delta} \bm{z}^{(k)},
\end{align}
where $\bm{z}^{(k)} \sim \mathcal{N}(\bm{0}, \bm{I}_N)$ is standard Gaussian noise.
By repeating this update for $K$ steps (possibly with a decaying step size schedule), the sequence approaches the target posterior distribution.
Since the first term inside the parenthesis is computed efficiently via \ac{AutoDiff} as described in \eqref{eq:vjp}, the entire sampling process remains computationally tractable.
We thus can generate samples $\bm{x}^{(k)} \sim p(\bm x|\bm x_{\text{in}}, \bm y)$ and approximate 
\begin{equation}
    \bm{x}_{post} = \mathbb{E}[\bm{X} | \bm{x}_{in}, \bm{y}]  \approx \frac{1}{K} \sum_{k=1}^K \bm{x}^{(k)},
\end{equation}
where $K$ is the number of samples.
The estimated value of $\bm{x}_{post}$ can be used in the \ac{SISO} module instead of \eqref{eq:tweedie}.

This hybrid approach allows the \ac{SC-VAMP} algorithm to leverage the exact physical model $f(\bm{x})$ without training a separate score network for the likelihood, while still utilizing the powerful learned score $s_{\theta}(\bm{x})$ for the complex prior in Module B.

\subsection{Computational Complexity}

One potential challenge of the proposed hybrid approach is the computational cost associated with the iterative Langevin sampling, especially when the forward operator $f(\bm{x})$ involves heavy numerical simulations (\textit{e.g.}, solving \acp{PDE} as in \cite{wadayama-2025-1}).
However, this cost can be significantly mitigated by leveraging the iterative nature of \ac{SC-VAMP}.
Since the estimate $\bm{x}_{in}$ changes gradually across \ac{SC-VAMP} iterations, we can employ a \textit{warm-start} strategy, initializing the Langevin chain with the current input $\bm{x}_{in}$ or the samples from the previous iteration. 
This strategy drastically reduces the number of steps $K$ required for burn-in.
Furthermore, the sampling process is inherently parallelizable.
By evaluating $f(\bm{x})$ for multiple particles simultaneously on a \ac{GPU}, the wall-clock time can be kept within a reasonable range for many practical applications.
Ultimately, this approach offers a trade-off: it incurs a higher computational cost compared to simple linearization (\textit{e.g.}, Extended \ac{VAMP}) but provides a more accurate score estimation in strongly nonlinear regimes where linearization fails.

\end{document}